%% file: main_v3.tex
\documentclass[journal,twocolumn,romanappendices]{IEEEtran}
\IEEEoverridecommandlockouts

\usepackage{amsmath,amssymb,amsthm}
\usepackage{bbm}

\theoremstyle{remark}

\usepackage{graphicx}
\graphicspath{{Figure/}}

\usepackage[ruled,vlined,linesnumbered]{algorithm2e}
\SetKwInput{KwInput}{Input}
\SetKwInput{KwOutput}{Output}

\usepackage{tikz}
\usepackage{pgfplots}
\usepgfplotslibrary{groupplots}
\pgfplotsset{compat=1.18}
\pgfplotsset{table/search path={.,files_tex}}
\usepackage{siunitx}

\usepackage{microtype}
\usepackage{url}
\usepackage{xspace}
\usepackage{soul}
\usepackage{enumitem}
\usepackage{setspace}
\usepackage{cite}
\usepackage[utf8]{inputenc}
\usepackage[T1]{fontenc}
\usepackage{amsfonts}

\usepackage[nolist]{acronym}

\usepackage[USenglish]{babel}
\usepackage[T1]{fontenc}
\usepackage[utf8]{inputenc}

\usepackage{subcaption}

\usepackage[figurename=Fig.,labelsep=period,font=small]{caption}

\usepackage{hyperref}

\usepackage[colorinlistoftodos]{todonotes}
\usepackage{pdfcomment}

\usepackage{xcolor}

\usepackage{titlesec}
\let\subparagraph\relax
\titlespacing{\section}{0pt}{6pt plus 2pt minus 1pt}{4pt plus 1pt minus 1pt} 
\titlespacing{\subsection}{0pt}{4pt plus 2pt minus 1pt}{2pt plus 1pt minus 1pt} 
\input{./Definitions.tex}
\input{./Acronyms.tex}

\title{Blind Channel Estimation and Data Detection for Near-Field XL-MIMO Systems}
\author{Maral Safari and Italo Atzeni
\thanks{The authors are with the Centre for Wireless Communications, University of Oulu, Finland (e-mail: \{maral.safarinaij, italo.atzeni\}@oulu.fi).}
\thanks{This work was supported by the Research Council of Finland (336449 Profi6, 348396 HIGH-6G, and 369116 6G Flagship).}}

\begin{document}

\maketitle

\begin{abstract}
Future wireless systems are expected to employ extremely large-scale multiple-input multiple-output (XL-MIMO) arrays at high carrier frequencies, where near-field propagation makes the channel depend jointly on angle and distance. The resulting short coherence intervals make channel state information acquisition challenging, motivating blind channel estimation and data detection (B-CE-DD). In this paper, we propose a two-stage B-CE-DD framework for uplink near-field XL-MIMO systems. First, we formulate the problem as the recovery of user-specific rank-one channel-data products from a superimposed received signal using a polar-domain sparse channel model and a low-dimensional data subspace model. Building on this formulation, we develop an on-grid blind orthogonal matching pursuit (B-OMP) algorithm that exploits polar-domain sparsity to iteratively identify the dominant angle-distance components and estimate the corresponding channel-data products, followed by an off-grid refinement stage based on block-coordinate descent (BCD) that optimizes the angle and distance parameters in the continuous polar domain. Numerical results show that the proposed B-CE-DD framework combining B-OMP and BCD significantly improves the symbol error rate compared with a pilot-based baseline employing zero-forcing beamforming, particularly at low signal-to-noise ratio and when the number of data symbols is small relative to the length of the coherence interval.
\end{abstract}


\section{Introduction} \label{sec:Intro}

Future wireless systems are envisioned to deliver extremely high data rates and support massive numbers of connected devices~\cite{rajatheva2020white}. Achieving these goals will require operation at higher carrier frequencies, from the upper mid-band to the \ac{sub-THz} range~\cite{petrov2020ieee,Atz25}, together with the deployment of large antenna arrays~\cite{chowdhury20206g}. While these frequencies provide wide bandwidths, their severe path loss and blockage sensitivity make large arrays indispensable, pushing conventional massive \ac{MIMO} toward the \ac{XL-MIMO} regime. In this context, propagation primarily occurs in the near field, where the large array size and high carrier frequency make the channel depend on both the angle and distance between the transmitter and receiver, rather than on the angle alone as in the far-field case~\cite{cui2022channel, zhang20236g}. 

At such high carrier frequencies and large array apertures, the channel becomes highly time-varying~\cite{chaccour2022seven} and spatially non-stationary~\cite{de2020non,lu2024nearfieldxltutorial}, which results in shorter coherence blocks and limits the accuracy of the \ac{CSI}. Under these conditions, traditional pilot-based channel estimation becomes inefficient, since accurate \ac{CSI} acquisition requires a large number of pilot symbols within each coherence block~\cite{ngo2013energy}. \textit{\Ac{B-CE-DD}} has the potential to address this issue by jointly recovering the channel and transmitted data directly from the received signal, without requiring dedicated pilot symbols. By exploiting the unknown data symbols as part of the estimation process, blind and semi-blind approaches can significantly reduce the training overhead or improve the detection performance for a given training overhead, which is particularly beneficial in presence of short coherence blocks~\cite{singh2019semi, razavikia2023off}.

\subsection{Prior Works and Motivation}

Most works on near-field \ac{XL-MIMO} channel estimation combine sparse recovery techniques with spherical-wavefront modeling. In~\cite{cui2022channel}, a pilot-based parametric framework is proposed, where a polar-domain transformation matrix captures the angle-distance dependency of near-field propagation. A method based on \ac{OMP} is employed to obtain on-grid estimates of the channel parameters, followed by a refinement stage based on \ac{GD} to alleviate grid mismatch effects. Building on this framework,~\cite{zhang2023near} introduces a distance-parameterized sparse representation and a dictionary-learning-based \ac{OMP} approach, which improves coherence properties while significantly reducing the dictionary size. In~\cite{lu2023mixed}, a two-stage channel estimation scheme is proposed for mixed \ac{LoS} and \ac{NLoS} near-field \ac{XL-MIMO} systems, which first estimates and removes the dominant \ac{LoS} path and then reconstructs the remaining \ac{NLoS} components using \ac{OMP} in the polar domain. The performance limits of parametric near-field channel estimation are analyzed in~\cite{gurgunouglu2025performance} using a two-dimensional MUSIC algorithm for joint angle-distance estimation.

Near-field propagation in \ac{XL-MIMO} also reshapes beamforming: instead of directional steering, energy can be focused at specific spatial locations, motivating new beamforming designs. The work in~\cite{bjornson2021primer} provides foundational insights through an analytical study of spherical-wave propagation and finite-depth focusing, characterizing when distance-dependent beam focusing is achievable and how the beamforming gain transitions between near- and far-field regimes. Building on these insights,~\cite{zhang20236g} describes the transition from far-field beam steering to near-field beam focusing and outlines key design challenges and opportunities. In~\cite{hu2023design}, a three-dimensional beamforming scheme for large intelligent surfaces is proposed, which accounts for spherical-wave propagation and distance-dependent phase variations to enable accurate spatial focusing in the Fresnel region. Furthermore,~\cite{kosasih2025near} develops a scalable modular subarray architecture that enlarges the effective aperture while reducing complexity and enabling distance-based spatial multiplexing.

Due to the short coherence blocks at high carrier frequencies, pilot-based channel estimation schemes can incur severe training overhead. This motivates blind or semi-blind approaches that aim to recover the channel and transmitted data without relying on dedicated pilots. In general, problems involving the recovery of multiple unknown signals from bilinear measurements arise in many signal processing applications and are commonly formulated as blind bilinear inverse problems. Blind deconvolution, i.e., the problem of recovering two unknown signals from their convolution, is investigated in~\cite{ahmed2013blind} when the signals lie in known low-dimensional subspaces and in~\cite{chi2016guaranteed} when one signal is sparse while the other lies in a known low-dimensional subspace. Furthermore,~\cite{li2019rapid} provides a non-convex formulation of blind deconvolution and solves it using a \ac{GD}-based algorithm. In a \ac{MIMO} system, the received signal depends bilinearly on the unknown channel and transmitted data vectors, and can be expressed as a superposition of rank-one channel-data products. Recovering both quantities from the received signal therefore leads to a structured blind bilinear inverse problem, which is closely related to blind deconvolution. Consequently, \ac{B-CE-DD} in \ac{MIMO} systems can be naturally cast within the blind bilinear inverse problem framework, an idea explored in several works.

For instance,~\cite{neumann2015channel} reviews massive \ac{MIMO} channel estimation and highlights the potential of blind and semi-blind approaches based on subspace and statistical properties. In~\cite{gao2007blind}, a \ac{B-CE-DD} scheme for \ac{MIMO} systems with orthogonal frequency-division multiplexing is proposed based on nonredundant linear precoding, enabling pilot-free channel estimation. In~\cite{mezghani2017blind}, the blind estimation of massive \ac{MIMO} channels is formulated as a maximum-likelihood problem exploiting angular-domain sparsity, while~\cite{Lassami2021} develops a \ac{B-CE-DD} scheme based on regularized deterministic maximum likelihood. Moreover,~\cite{vargas2023dual} investigates dual-blind deconvolution for integrated radar-communication systems. The work in~\cite{bigdeli2022noncoherent} builds on the super-resolution framework and proposes an off-grid \ac{B-CE-DD} method based on atomic norm minimization~\cite{candes2014towards,safari2021off}. Lastly,~\cite{arai2025joint} proposes a pilot-assisted channel and data estimation framework for \ac{XL-MIMO} systems based on non-orthogonal pilot transmission, which first performs an initial channel estimation and subsequently iterates between data estimation and residual channel-error refinement through a probabilistic inference procedure.

Blind approaches have strong potential to improve the detection performance in \ac{XL-MIMO} systems. However, near-field \ac{B-CE-DD} cannot be achieved through a straightforward extension of conventional far-field methods, nor by merely replacing the angular dictionary with a polar-domain one. Existing blind approaches for \ac{MIMO} systems rely not only on sparsity, but also on structural properties induced by planar-wave propagation, such as angle-only steering vectors and the associated subspace structure. Moreover, they often rely on orthogonal pilot or precoding schemes that enable user separation, or on standard linear or convolutional models. In contrast, we consider a near-field setting in which the channel depends jointly on angle and distance and the received signal is a superposition of all users’ contributions, with the unknowns being user-specific rank-one channel-data products. This leads to a fundamentally different structured bilinear inverse problem with inherent inter-user coupling.

\subsection{Contributions}

In this paper, we propose a two-stage \ac{B-CE-DD} framework for near-field \ac{XL-MIMO} systems, comprising an on-grid stage based on \ac{OMP} and a subsequent off-grid refinement stage based on \ac{BCD}. The main contributions are summarized as follows.
\begin{itemize}
\item Considering an uplink near-field \ac{XL-MIMO} system, we formulate the \ac{B-CE-DD} problem as the recovery of user-specific rank-one channel-data products from a superimposed received signal by combining a polar-domain sparse channel model with a low-dimensional data subspace model, which yields a bilinear representation. User-specific processing without inter-user interference can be achieved when the total number of data symbols transmitted across all the users is smaller than the length of the coherence interval. This regime is practically relevant when each user transmits only a limited amount of data within each coherence interval, as in short-packet transmissions. In this setting, jointly recovering the channel and transmitted data can be more effective than treating channel estimation and data detection as separate tasks, since the received signal is exploited more efficiently through their coupled structure.
 
\item We develop the \textit{\ac{B-OMP}} algorithm, which jointly estimates the near-field channel and transmitted data vectors under an on-grid model. \ac{B-OMP} leverages the fact that each user's channel vector can be accurately represented by a few atoms, each corresponding to a steering vector associated with a candidate angle-distance pair from a predefined polar-domain dictionary. In this setting, \ac{B-OMP} iteratively selects the atoms corresponding to the dominant angle-distance pairs and updates the estimates of the channel-data products. After convergence, the estimated channel-data products are factorized and phase-aligned to recover the data vectors.

\item We develop a \ac{BCD} algorithm to mitigate the approximation error arising from the finite grid resolution of the on-grid stage. In this refinement stage, the angle and distance estimates are no longer restricted to predefined polar-domain grid points. For each user, the proposed method minimizes the residual error between the effective received signal and its reconstruction based on the channel and data estimates, updating the estimated polar-domain channel parameters (i.e., angles, distances, and small-scale fading coefficients) and the data vector in a \ac{BCD} fashion. The proposed \ac{BCD} algorithm converges to a stationary point of the objective function.

\item We analyze the computational complexity of the proposed \ac{B-OMP} and \ac{BCD} algorithms. The complexity of \ac{B-OMP} scales linearly with the number of antennas, length of the coherence interval, and size of the polar-domain dictionary, whereas the complexity of \ac{BCD} grows quadratically with the number of antennas.

\item We demonstrate through numerical results that, when the total number of data symbols transmitted across all the users is smaller than the length of the coherence interval, the proposed near-field \ac{B-CE-DD} framework achieves substantial performance gains over a baseline employing \ac{ZF} beamforming with channel estimates obtained with the \ac{OMP}-based method from~\cite{cui2022channel}. In particular, we evaluate the \ac{SER} under different configurations of \ac{SNR}, number of data symbols relative to the length of the coherence interval, channel sparsity, and modulation order. The gains are especially pronounced when the per-antenna, per-symbol \ac{SNR} is low and the number of data symbols is small compared with the length of the coherence interval. 
\end{itemize}

\subsection{Notations and Paper Organization}

Boldface lowercase and uppercase letters represent vectors and matrices, respectively, whereas calligraphic letters denote sets. $(\cdot)^{\tran}$, $(\cdot)^{\ast}$, and $(\cdot)^{\herm}$ denote the transpose, conjugate, and conjugate transpose operators, respectively. Considering $\A \in \Compl^{m \times n}$, $\A(p,q) \in \Compl$ denotes the $(p,q)$-th entry of $\A$, whereas $\A(:,q_1:q_2) \in \Compl^{m \times (q_2-q_1+1)}$ represents the sub-matrix obtained from $\A$ by retaining the columns from $q_1$ to $q_2$; 
further considering the index set $\setI$, $\A(\setI,:) \in \Compl^{|\setI| \times n}$ and $\A(:,\setI) \in \Compl^{m \times |\setI|}$ represent the sub-matrices obtained from $\A$ by retaining the rows and columns indexed by $\setI$, respectively. $[\a_{1}, \ldots, \a_{N}]$ and $[\A_{1}, \ldots, \A_{N}]$ denote horizontal concatenation of the vector and matrix arguments, respectively, whereas $\blkdiag (\A_{1}, \ldots, \A_{N})$ produces a block-diagonal matrix of the matrix arguments. The pseudoinverse of $\A$ (tall matrix with full column rank) is denoted by $\A^{\dagger} = (\A^\herm\A)^{-1}\A^\herm$. $\| \cdot \|_{2}$ and $\| \cdot \|_\mathrm{F}$ represent the Euclidean norm for vectors and the Frobenius norm for matrices, respectively, whereas $\tr(\cdot)$ denotes the trace operator. For a function $f(\x)$ with $\x = [x_1,\ldots,x_n]^\tran$, $\nabla_{\x}f(\x)$ and $\frac{\partial f(\x)}{\partial x_i}$ denote the gradient with respect to $\x$ and the partial derivative with respect to $x_i$, respectively. The circularly symmetric complex normal distribution with zero mean and variance $x^2$ is denoted by $\mathcal{CN}(0, x^2)$, whereas the uniform distribution over the interval $[x_{\textrm{min}}, x_{\textrm{max}}]$ is denoted by $\mathcal{U} [x_{\textrm{min}}, x_{\textrm{max}}]$. $\Re[\cdot]$ represents the real part and $j = \sqrt{-1}$ the imaginary unit.

The remainder of this paper is organized as follows. Section~\ref{system model} introduces the system model and formulates the considered \ac{B-CE-DD} problem. Section~\ref{main section} presents the proposed near-field \ac{B-CE-DD} framework, which is our core contribution: specifically, Section~\ref{on grid} describes the on-grid stage based on the \ac{B-OMP} algorithm, while Section~\ref{off grid} illustrates the off-grid refinement stage based on the \ac{BCD} algorithm. Section~\ref{simulation} reports numerical results comparing the proposed framework with a conventional pilot-based scheme. Finally, Section~\ref{sec:concl} concludes the paper. Some detailed derivations are provided in the appendices.

\section{System Model}\label{system model}

We consider an uplink \ac{XL-MIMO} system where $K$ single-antenna users simultaneously transmit data to a \ac{BS} equipped with $N$ antennas. Let $\h_{k} \in \Compl^{N}$ be the channel vector from the $k$-th user to the \ac{BS}, with  $k \in \{1,\dots,K\}$. The overall channel matrix is defined as  $\H = [\h_{1},\ldots, \h_{K}] \in \Compl^{N \times K}$ and is assumed to remain constant within a coherence interval of $T$ symbols. The $k$-th user transmits the signal $\x_{k} \in  \Compl^{T}$. Then, the received signal at the \ac{BS} is given by
\begin{align}\label{main_in_out}
   \Y = \sqrt{\rho}\sum^{K}_{k = 1}  \h_{k}\x_{k}^{\tran} + \Z \in \Compl^{N \times T},
\end{align} 
where $\rho$ is the transmit power (assumed equal for all the users) and $\Z$ is the \ac{AWGN} matrix with independent $\mathcal{CN}(0,\sigma^2)$ entries, where $\sigma^{2}$ is the \ac{AWGN} variance. 

In the following, we present the near-field channel model used throughout the the paper, incorporating both angle- and distance-dependent characteristics.

\subsection{Channel Model}

We assume that the \ac{BS} is equipped with a \ac{ULA}, with inter-antenna spacing denoted by $d$, and consider near-field propagation based on spherical wavefronts. Following the channel model in~\cite{cui2022channel}, the channel vector of the $k$-th user is expressed as
\begin{align}\label{near-field channel}
    \h_{k}=\frac{1}{\sqrt{L}}\sum^{L}_{l = 1} g_{k,l}\b(\theta_{k,l},r_{k,l}) \in \Compl^{N},
\end{align}
where $L$ is the number of propagation paths that determines the channel sparsity, $g_{k,l} \in \Compl$ denotes the small-scale fading coefficient associated with the $l$-th path of the $k$-th user, and $\theta_{k,l} \in \big[-\frac{\pi}{2}, \frac{\pi}{2}\big]$ and $r_{k,l} \in \Real_{+}$ represent, respectively, the angle (with respect to the \ac{ULA}'s broadside direction) and distance corresponding to the $l$-th path of the $k$-th user, both referenced to the center of the \ac{ULA}. Furthermore, $\b(\theta_{k,l},r_{k,l}) \in \Compl^{N}$ is the near-field steering vector defined as
\begin{align}\label{near-field steering vector}
   \b(\theta_{k,l},r_{k,l}) = [e^{-j\frac{2 \pi}{\lambda_{\textrm{c}}} (r^{(0)}_{k,l}-r_{k,l})},\ldots, e^{-j\frac{2 \pi}{\lambda_{\textrm{c}}} (r^{(N-1)}_{k,l}-r_{k,l})}]^\tran,
\end{align}
where 
\begin{align}
r^{(n)}_{k,l} = \sqrt{r^{2}_{k,l} + n^2d^2 - 2r_{k,l}n d\sin\theta_{k,l}}, \, \forall n \in \{0, \ldots, N -1 \}
\end{align}
is the distance corresponding to the $l$-th path of the $k$-th user referenced to the $n$-th antenna and $\lambda_\textrm{c}$ denotes the wavelength corresponding to the adopted carrier frequency. Unlike conventional far-field models, which depend solely on angular information of the users/scatterers, near-field channels are jointly determined by both angles and distances. This distinction stems from the spherical-wavefront setting, under which the propagation path length of the path indexed by $(k,l)$ varies across antennas as a function of both $\theta_{k,l}$ and $r_{k,l}$. Hence, $\h_{k}$ can be expressed as a weighted sum of near-field steering vectors, each associated with a distinct angle-distance pair.

Because only a few dominant propagation paths exist (originating from a few strong scatterers and possibly from the \ac{LoS} link), only a small subset of angle-distance pairs contributes significantly to the received signal. This observation motivates representing the channel in the polar domain. Let $\W \in \Compl^{N\times Q}$ denote the polar-domain transformation matrix, whose $Q$ columns consist of steering vectors associated with predefined angle-distance pairs~\cite{cui2022channel}. Then, the channel vector of the $k$-th user can be approximated as\footnote{The approximation accounts for the fact that the actual propagation paths may not align exactly with the angle-distance pairs used to construct $\W$.}
\begin{align} \label{eq:Wg_k}
     \h_{k} \approx \W\g_k ,
\end{align}
where $\g_k \in \Compl^{Q}$ is the polar-domain channel vector. Stacking all the users, we define $\G = [\g_1,\ldots,\g_K] \in \Compl^{Q \times K}$. In this representation, $\g_k$ is a compressible vector containing mostly negligible entries, while its few significant components indicate the dominant propagation paths of the $k$-th user in the polar domain. This inherent polar-domain sparsity will be exploited in Section~\ref{main section} to model and process the received signal for \ac{B-CE-DD}.

\subsection{Received Signal}

\Ac{B-CE-DD} is generally an ill-posed problem, which may not have a unique or reliable solution unless additional structural constraints are imposed. To make the problem tractable, we assume that each user's transmitted signal lies in a known low-dimensional subspace~\cite{ahmed2013blind,ling2019regularized,chi2016guaranteed, bigdeli2022noncoherent}. For each user~$k$, let $\d_k \in \Compl^{S}$ denote the data vector to be recovered, with $S \ll T$ being the number of data symbols. The data symbols are drawn from a complex modulation constellation; in our numerical results in Section~\ref{simulation}, we consider \ac{QAM} constellations with different modulation orders $M$. We introduce the precoding matrix $\C_k \in \Compl^{T \times S}$, applied at the user but also known at the \ac{BS}, and model the transmitted signal as $\x_{k} = \C_k \d_k$. Hence, the received signal in \eqref{main_in_out} can be rewritten as
\begin{align}\label{in-out-in-the-without_pilot_symbol}
    \Y  = \sqrt{\rho}\sum^K_{k=1}\h_k\d_k^{\tran}\C_k^{\tran} + \Z.
\end{align}
The goal of \ac{B-CE-DD} is to estimate the channel-data product $\h_k\d_k^{\tran}$ and then factorize it to recover the data vector $\d_k$, $\forall k \in \{1, \ldots, K\}$.

Without prior information, the separation of the channel and data vectors is ambiguous up to a complex scalar. This induces: \textit{i)} a phase ambiguity, which is removed by introducing a known pilot symbol for each user; and \textit{ii)} an amplitude ambiguity, which is resolved by fixing the norm of each data vector to a known value. For each user~$k$, let $p_k \in \Compl$ denote the pilot symbol and define the augmented data vector $\bar{\d}_k = [p_k, \d_k^\tran]^\tran \in \Compl^{S+1}$. The pilot symbol provides a phase reference that allows to estimate and correct the user-specific phase ambiguity, enabling reliable detection of complex modulation constellations~\cite{zeng2004semi,gao2007blind}. Furthermore, we impose the unit-norm constraint $\|\bar{\d}_k\|_2 = 1$, which eliminates the amplitude ambiguity since the magnitude of the augmented data vector is known. Let $\bar{\C}_k \in \Compl^{T\times(S+1)}$ denote the augmented precoding matrix. The augmented received signal incorporating the user-specific pilot symbols is given by (cf. \eqref{in-out-in-the-without_pilot_symbol})
\begin{align} \label{in-out-in-the-with_pilot_symbol}
\begin{split}
    \bar{\Y} & = \sqrt{\rho}\sum^K_{k=1}\h_k\bar{\d}_k^{\tran}\bar{\C}_k^{\tran} + \Z \\
    & = \sqrt{\rho}\sum^K_{k=1}\W\underbrace{\g_k\bar{\d}_k^{\tran}}_{=\boldsymbol{\Xi}_k}\bar{\C}_k^{\tran} + \Z \in \Compl^{N \times T},
\end{split}
\end{align}
where the second equality follows from utilizing the channel representation in \eqref{eq:Wg_k}. Thus, the bilinear representation in \eqref{in-out-in-the-with_pilot_symbol} consists of a superposition of linear transformations of the rank-one matrices \(\{\boldsymbol{\Xi}_k\}_{k=1}^{K}\), where $\boldsymbol{\Xi}_k = \g_k\bar{\d}_k^{\tran}\in \Compl^{Q\times (S+1)}$ represents the channel-data product of the $k$-th user. Now, defining $\bar{\C} = \blkdiag (\bar{\C}_1, \ldots, \bar{\C}_K)\in \Compl^{TK\times K(S+1)}$, $\bar{\D} = \blkdiag (\bar{\d}_1, \ldots, \bar{\d}_K) \in \Compl^{K(S+1)\times K}$, $\boldsymbol{\Xi} = [\boldsymbol{\Xi}_1, \ldots, \boldsymbol{\Xi}_K]\in \Compl^{Q \times K(S+1)}$, and $\A =\1^{\tran}_K\otimes \I_T \in \Real^{T \times T K}$, we rewrite \eqref{in-out-in-the-with_pilot_symbol} as
\begin{align} \label{input-output_rewrite}
\begin{split}
    \bar{\Y} & = \sqrt{\rho} \W\G \bar{\D}^\tran \bar{\C}^\tran\A^\tran + \Z \\
    & = \sqrt{\rho}\W\boldsymbol{\Xi}\bar{\C}^\tran\A^\tran + \Z,
\end{split}
\end{align}
which serves as the input to our \ac{B-CE-DD} framework in Section~\ref{main section}.

The \ac{OMP}-based channel estimation method in~\cite{cui2022channel} relies on orthogonal pilot resources across the $K$ users, enabling channel estimation to be performed independently for each user. Similarly, the channel and data estimation framework in~\cite{arai2025joint} first obtains an initial channel estimate from dedicated pilot observations and subsequently refines the channel through iterative updates based on the detected data symbols. In contrast, the proposed approach aims to jointly recover the channel and transmitted data simultaneously from the superimposed received signal in \eqref{input-output_rewrite} without relying on a dedicated pilot-based channel estimation stage. Since the received signals from all the users are superimposed in the considered setting, user-specific processing from the aggregate received signal in \eqref{input-output_rewrite} without inter-user interference requires the composite matrix $\A\bar{\C} \in \Compl^{T\times K(S+1)}$ to have full column rank or, equivalently, $\bar{\C}^\tran\A^\tran$ to be left-invertible. A sufficient condition for this requirement is $T \ge K(S{+}1) \approx KS$, which ensures that the signals of different users can be effectively separated. This regime is practically relevant when each user transmits only a limited amount of data within each coherence interval, as in short-packet transmissions. In this context, the number of users that can be multiplexed grows approximately linearly with the ratio $\frac{T}{S}$.

Next, we propose a near-field \ac{B-CE-DD} framework that integrates on-grid and off-grid channel estimation approaches: first, the channel-data product for each user is estimated using an \ac{OMP}-based method, after which the data vectors are recovered (on-grid stage); then, the channel and data estimates are refined via \ac{BCD} (off-grid refinement stage).

\section{Proposed Near-Field \ac{B-CE-DD} Framework} \label{main section}

In this section, we develop a two-stage \ac{B-CE-DD} framework for near-field \ac{XL-MIMO} systems, which includes an on-grid stage based on \ac{OMP} followed by an off-grid refinement stage based on \ac{BCD}. Section~\ref{on grid} focuses on the on-grid stage, where the estimated channel parameters are confined to a predefined polar-domain grid. For this first stage, we propose a sparse recovery method termed \textit{\acf{B-OMP}}, which first estimates the channel-data product for each user and then extracts the data vector by means of \ac{SVD} followed by a phase alignment. Subsequently, Section~\ref{off grid} addresses the off-grid refinement stage, where the estimated channel parameters can deviate from the polar-domain grid points. This second stage is initialized with the channel and data estimates obtained via \ac{B-OMP}, which are refined in the continuous polar domain using \ac{BCD}.

\subsection{On-Grid \ac{B-CE-DD}: \ac{B-OMP} Algorithm}\label{on grid}

In this section, we introduce the proposed \ac{B-OMP} algorithm, which jointly estimates the near-field channel and data vectors under an on-grid model. Unlike traditional pilot-based schemes, \ac{B-OMP} operates without explicit training sequences, except for a single pilot symbol per user introduced to resolve the inherent phase ambiguity. The proposed scheme exploits the sparsity of near-field channels in the polar domain, where each user's channel vector can be accurately represented by a small number of atoms, each corresponding to a steering vector associated with a candidate angle-distance pair from a predefined polar-domain dictionary. Leveraging this structure, \ac{B-OMP} iteratively selects the atoms corresponding to the dominant angle-distance pairs and updates the estimates of the channel-data products. After convergence, the estimated channel-data products are factorized to recover the data vectors. The key distinction from~\cite{cui2022channel} is in the estimation objective: while~\cite{cui2022channel} focuses solely on channel estimation, the proposed \ac{B-OMP} first estimates the channel-data product and subsequently separates it to obtain the channel and data vectors. Hence, the different estimation objective leads to a substantially different algorithmic procedure, which is summarized in Algorithm~\ref{algo1} and described in detail below.

\begin{algorithm}[t!]
\small
\SetAlgoLined
\setlength{\AlCapSkip}{0.75em}
\caption{On-grid \ac{B-CE-DD} via \ac{B-OMP}}
\label{algo1}

\KwInput{Received signal $\bar{\Y}$; number of paths $L$ or threshold $\epsilon_1$\;}\nllabel{algo1:input}

\textbf{Initialization:} Set $\R^{(0)} = \bar{\Y}$ and $\mathcal{S}_{k}^{(0)} = \emptyset$, $\forall k \in \{ 1, \ldots, K \}$; set $i=0$\;\nllabel{algo1:init}

\Repeat{$i = L$ \ \textbf{\textup{or}} \ $\|\bar{\Y} - \R^{(i)}\|_{\mathrm{F}}^{2} \le \epsilon_1$}{%

    \nllabel{algo1:forL-start}%
    Set $i=i+1$\;
    Compute the correlation matrix $\boldsymbol{\Gamma}^{(i)}$ as in \eqref{correlation matrix}\;\nllabel{algo1:corr}

    \For{$k = 1, \ldots, K$}{%
        \nllabel{algo1:forK-start}%
        Extract the user-specific correlation matrix $\boldsymbol{\Gamma}_k^{(i)}$ as in \eqref{gamma_k}\;\nllabel{algo1:gamma}
        Select the index $s_k^{(i)}$ of the propagation path that is most correlated with the current residual as in \eqref{energy}\;\nllabel{algo1:energy}
        Update the support set $\mathcal{S}^{(i)}_k$ as in \eqref{eq:support}\;\nllabel{algo1:support}
        Compute the estimate of the channel-data product $\hat{\boldsymbol{\Xi}}_{k}^{(i)}$ as in \eqref{least-square-algo1}\;\nllabel{algo1:proj}
    }

    Update the residual $\R^{(i)}$ as in \eqref{residual_algo1}\;\nllabel{algo1:resid}
}
Set $\hat{\boldsymbol{\Xi}} = \hat{\boldsymbol{\Xi}}^{(i)}$\;\nllabel{algo1:set}
\For{$k = 1, \ldots, K$}{%
Obtain preliminary estimates $\hat{\g}_k$ and $\hat{\bar{\d}}_k$ as in \eqref{before phase correction_channel}--\eqref{before phase correction_data}\;\nllabel{algo1:prephase}

Apply phase alignment as in \eqref{Phase_alignment_rank_one}--\eqref{final_estimation_rank_one}\;\nllabel{algo1:phase}
}

\KwOutput{$\hat{\mathbf{G}} = [\hat{\mathbf{g}}_1, \ldots, \hat{\mathbf{g}}_K]$, $\hat\H = \W\hat\G$, and $\hat{\mathbf{D}} = [\hat{\mathbf{d}}_1, \ldots, \hat{\mathbf{d}}_K]$.}\nllabel{algo1:output}
\end{algorithm}

Starting from the observed received signal $\bar\Y$ in \eqref{input-output_rewrite}, we first obtain an estimate of the channel-data product $\boldsymbol{\Xi}_k = \g_{k} \bar{\d}_{k}^{\tran}$ for each user~$k$, denoted by $\hat{\boldsymbol{\Xi}}_k \in \Compl^{Q \times (S+1)}$, via an \ac{OMP}-based method, as detailed in Section~\ref{sub_sub_OMP}. Since $\boldsymbol{\Xi}_k$ is a rank-one matrix, the estimated polar-domain channel vector $\hat{\g}_{k} \in \Compl^{Q}$ and the estimated data vector $\hat{\bar{\d}}_{k} \in \Compl^{S+1}$ are obtained as the principal left and right singular vectors of $\hat{\boldsymbol{\Xi}}_k$, respectively. This step, described in Section~\ref{sub_sub_SVD}, can be carried out via a standard \ac{SVD} or a lower-complexity power iteration method, followed by a phase alignment. The procedure is aided by the user-specific pilot symbol $p_{k}$ and the unit-norm constraint $\|\bar{\d}_k\|_2 = 1$.

\smallskip

\subsubsection{Estimation of the Channel-Data Products}\label{sub_sub_OMP}

In a general sparse recovery problem, an \ac{OMP}-based algorithm iteratively selects the dictionary atoms most correlated with the current residual, expands the active support set, re-estimates the signal over the selected atoms, and updates the residual, until a predefined termination criterion is satisfied. To estimate the channel-data products in our near-field \ac{B-CE-DD} framework, we define the residual $\R^{(i)} \in \Compl^{N \times T}$ representing the part of the received signal that remains after subtracting the contribution of the atoms (corresponding to steering vectors in $\W$) selected up to the $i$-th iteration. To track the estimated propagation paths for each user~$k$, we introduce the user-specific support set $\mathcal{S}^{(i)}_k$, which contains the indices of the atoms selected for that user up to the $i$-th iteration, with $|\mathcal{S}^{(i)}_k| = i$. Since no propagation paths have been estimated at the start of the algorithm, the residual and support sets are initialized as $\R^{(0)} = \bar{\Y}$ and $\mathcal{S}_{k}^{(0)} = \emptyset$, $\forall k \in \{ 1, \ldots, K \}$, respectively (see line~\ref{algo1:init}).

Let us define the effective received signal, i.e., the received signal after removing the effect of the precoding matrix, as
\begin{align} \label{eq:Y_breve}
\begin{split}
\Breve{\Y} 
&= \bar{\Y}(\bar\C^\tran\A^\tran)^{\dag} \\
&= \sqrt{\rho}\W \boldsymbol{\Xi} + \breve{\Z} \\
&= [\Breve{\Y}_1, \ldots, \Breve{\Y}_K] \in \Compl^{N \times K(S+1)},
\end{split}
\end{align}
with
\begin{align}
\Breve{\Y}_k = \sqrt{\rho}\W \boldsymbol{\Xi}_k + \breve{\Z}_k \in \Compl^{N \times (S+1)} \label{Yk_block}
\end{align}
and $\breve{\Z} = \Z(\bar\C^\tran\A^\tran)^{\dag} = [\breve{\Z}_1, \ldots, \breve{\Z}_K] \in \Compl^{N \times K(S+1)}$, where the last equality in \eqref{eq:Y_breve} follows from the user-specific block structure of the composite matrix $\A\bar{\C}$. Therefore, \eqref{eq:Y_breve} collects the user-specific effective received signals in the form of \eqref{Yk_block}, arranged in a concatenated manner. At each iteration~$i$, the column of $\W$ that is most correlated with the current residual $\R^{(i-1)}$ is selected. To this end, the correlation matrix
\begin{align}\label{correlation matrix} 
\boldsymbol{\Gamma}^{(i)} &= \W^\herm\R^{(i-1)}(\bar\C^\tran\A^\tran)^{\dag} \in \Compl^{Q\times K(S+1)}
\end{align}
is computed (see line~\ref{algo1:corr}). This matrix measures the correlation between each atom and the current residual, obtained from $\Breve{\Y}$ in \eqref{eq:Y_breve} after subtracting the contribution of the atoms selected in the previous iterations, and thus identifies the dominant propagation paths among those not yet selected. Similar to \eqref{eq:Y_breve}, the correlation matrix in \eqref{correlation matrix} can be partitioned into user-specific correlation matrices in the form of (see line \eqref{algo1:gamma})
\begin{align}\label{gamma_k}
\boldsymbol{\Gamma}^{(i)}_k 
= \boldsymbol{\Gamma}^{(i)}\big(:, (k - 1)(S + 1) + 1 : k(S + 1)\big)\in \Compl^{Q\times (S+1)}.
\end{align}
Since the subsequent processing can be carried out independently for each user, we focus on describing the procedure for the $k$-th user in the following.

The user-specific correlation matrix in \eqref{gamma_k} reflects the structural pattern of $\boldsymbol{\Xi}_k$, which is rank-one and whose columns 
are scaled versions of $\mathbf{g}_k$ (i.e., they share the same row support). 
Consequently, the support set $\mathcal{S}_k^{(i)}$ can be updated directly from $\boldsymbol{\Gamma}_k^{(i)}$ by identifying the index of the propagation path that is most correlated with the current residual. This is done by selecting the row of $\boldsymbol{\Gamma}_k^{(i)}$ with maximum power as (see line~\ref{algo1:energy})
\begin{align}\label{energy}
s_k^{(i)} = \argmax_{s} \sum_{m=1}^{S+1}\big|\boldsymbol{\Gamma}^{(i)}_k (s,m)\big|^2,
 \end{align}
which is used to update the support set as (see line~\ref{algo1:support})
\begin{align}\label{eq:support}
\mathcal{S}^{(i)}_k =\mathcal{S}^{(i-1)}_k \cup \{s_k^{(i)} \}.
\end{align} 
Let $\hat{\boldsymbol{\Xi}}^{(i)}_k \in \Compl^{Q \times (S+1)}$ denote the estimate of $\boldsymbol{\Xi}_k$ at the $i$-th iteration, consisting of $i$ non-zero rows indexed by $\mathcal{S}^{(i)}_k$. These rows are obtained by orthogonally projecting the effective received signal onto the subspace spanned by the 
columns of $\W(:, \mathcal{S}^{(i)}_k) \in \Compl^{N \times i}$. Specifically, we insert the matrix $\mathbf{W}(:,\mathcal{S}_{k}^{(i)})^{\dagger}\breve{\Y}_{k} \in \Compl^{i \times (S+1)} $ into the rows of $\hat{\boldsymbol{\Xi}}_{k}^{(i)}$ indexed by $\mathcal{S}_{k}^{(i)}$, i.e.,
\begin{align}\label{least-square-algo1}
\hat{\boldsymbol{\Xi}}_{k}^{(i)} (\mathcal{S}_{k}^{(i)},:)
= \mathbf{W}(:,\mathcal{S}_{k}^{(i)})^{\dagger} \breve{\mathbf{Y}}_{k},
  \end{align}
while the remaining rows of $\hat{\boldsymbol{\Xi}}_{k}^{(i)}$ are set to zero (see line~\ref{algo1:proj}). Then, we obtain the estimate of $\boldsymbol{\Xi}$ at the $i$-th iteration as $\hat{\boldsymbol{\Xi}}^{(i)} = [\hat{\boldsymbol{\Xi}}^{(i)}_1, \ldots, \hat{\boldsymbol{\Xi}}^{(i)}_K]\in\Compl^{Q\times K(S+1)}$, which contains the dominant propagation paths of all the users estimated up to the $i$-th iteration. Lastly, The residual is updated by subtracting the reconstructed contribution as (see line~\ref{algo1:resid})
\begin{align}\label{residual_algo1}
\R^{(i)} = \R^{(i-1)} - \W \hat{\boldsymbol{\Xi}}^{(i)}\bar{\C}^{\tran}\A^{\tran}.
\end{align}

The above procedure is repeated until a predefined termination criterion is satisfied (see line~\ref{algo1:resid}):
\begin{itemize}
\item If the number of propagation paths $L$ is known in advance, the algorithm can be configured to terminate after exactly $L$ iterations.
\item If the number of propagation paths $L$ is not known in advance, the algorithm can be configured to terminate when the residual error becomes sufficiently small, i.e., for $\|\bar{\Y} - \R^{(i)}\|_\mathrm{F}^{2} \leq \epsilon_1$, where $\epsilon_1$ is a predefined threshold. Since $\R^{(i)}$ represents the part of $\bar{\Y}$ that remains unexplained after the $i$-th iteration, ensuring that the residual error is negligible indicates that the dominant propagation paths have been estimated.
\end{itemize}
Finally, upon termination, we set $\hat{\boldsymbol{\Xi}} = \hat{\boldsymbol{\Xi}}^{(i)}$, with $i$ being the last iteration (see line~\ref{algo1:set}), and denote by $\hat{L}$ the number of propagation paths estimated by \ac{B-OMP}, which is equal to $L$ for the first termination criterion.

\smallskip

\subsubsection{Factorization via \ac{SVD} and Phase Alignment}\label{sub_sub_SVD}

To recover the channel and data vectors associated with each user~$k$ from the estimated channel-data product $\hat{\boldsymbol{\Xi}}_k$, the most straightforward approach is to apply \ac{SVD} as $\hat{\boldsymbol{\Xi}}_k = \U_k \boldsymbol{\Sigma}_k \V_k^{\herm}$, where \(\U_k \in \Compl^{Q\times Q}\) and \(\V_k\in \Compl^{(S+1)\times (S+1)}\) are unitary matrices containing the left and right singular vectors, respectively, and \(\boldsymbol{\Sigma}_k \in \Compl^{Q\times (S+1)}\) is a diagonal matrix containing the corresponding singular values. Let $\sigma_{k,1}$ denote the principal singular value for the $k$-th user, with corresponding left and right singular vectors \(\u_{k,1}\) and \(\v_{k,1}\), respectively. Then, we obtain preliminary estimates of the polar-domain channel and data vectors as (see line~\ref{algo1:prephase})
\begin{align}
    \tilde{\g}_k & = \sigma_{k,1} \u_{k,1} \in \Compl^{Q}, \label{before phase correction_channel}\\
    \tilde{\d}_k & = \v_{k,1} \in \Compl^{S+1},\label{before phase correction_data}
\end{align}
respectively. Note that $\sigma_{k,1}$ can be absorbed into $\tilde{\g}_k$ thanks to the unit-norm constraint assumed for the data vector. However, since an \ac{SVD}-based factorization is defined only up to a global phase ambiguity, $\tilde{\g}_k$ and $\tilde{\d}_k$ must be phase aligned.

The phase ambiguity is resolved by means of the known pilot symbol $p_k$ in the first entry of $\bar{\d}_k$. Let $\tilde{p}_k \in \Compl$ denote the corresponding entry of $\tilde{\d}_k$. The phase correction factor is given by
\begin{align}\label{Phase_alignment_rank_one} 
    \chi_k = \angle \bigg(\frac{p_{k}}{\tilde{p}_{k}} \bigg),
\end{align}
and the phase-aligned estimates are simply obtained as (see line~\ref{algo1:phase})
\begin{align}
    \hat{\g}_k &= e^{-j \chi_k} \tilde{\g}_k\in\Compl^{Q}, \\
    \hat{\bar{\d}}_k &= e^{j \chi_k} \tilde{\d}_k\in\Compl^{S+1}. \label{final_estimation_rank_one}
\end{align}
Finally, the estimate of $\d_{k}$, denoted by $\hat{\d}_{k} \in \Compl^{S}$, is obtained from $\hat{\bar{\mathbf{d}}}_k$ by removing its first entry. Stacking all the users yields $\hat{\mathbf{G}} = [\hat{\mathbf{g}}_1, \ldots, \hat{\mathbf{g}}_K]\in\Compl^{Q\times K}$ and $\hat{\mathbf{D}} = [\hat{\mathbf{d}}_1, \ldots, \hat{\mathbf{d}}_K]\in\Compl^{S\times K}$, with the estimated channel matrix reconstructed as $\hat\H = \W\hat\G = [\hat\h_1,\ldots,\hat\h_K] \in\Compl^{N\times K}$.

Since only the principal singular value and vectors are required in the above procedure, computing a full \ac{SVD} is unnecessary. A lower-complexity alternative is the \emph{power iteration method}, which estimates the principal singular components by iteratively multiplying an initial vector by $\hat{\boldsymbol{\Xi}}_k^{\herm}\hat{\boldsymbol{\Xi}}_k$ and normalizing after each step. The power iterations converge to the principal singular components and, since $\hat{\boldsymbol{\Xi}}_k$ is nearly rank-one, convergence is remarkably fast. This makes the power iteration method a simple and efficient substitute for a full \ac{SVD}.

Therefore, the proposed \ac{B-OMP} provides on-grid estimates of the support set, polar-domain channel, and data vector for all the users simultaneously, which together serve as the initialization for the off-grid refinement stage described next.

\subsection{Off-Grid \ac{B-CE-DD}: \ac{BCD} Algorithm}\label{off grid}

In this section, we describe the proposed \ac{BCD} algorithm, which mitigates the approximation error caused by the finite grid resolution of the on-grid stage in Section~\ref{on grid}. Unlike \ac{B-OMP}, where the angle and distance estimates were restricted to predefined polar-domain grid points, this refinement stage allows these parameters to take continuous values within the polar domain. The objective function is defined as the user-specific residual error between the effective received signal and its reconstruction based on the channel and data estimates. This objective is minimized via \ac{BCD} with respect to the estimated polar-domain channel parameters, namely angles, distances, and small-scale fading coefficients, as well as the data vector. Specifically, the angle and distance estimates are updated via \ac{GD}, whereas the estimated small-scale fading coefficients and data vector are updated in closed form through least-squares solutions. Unlike the off-grid method in~\cite{cui2022channel}, which refines only the polar-domain channel, the proposed method jointly updates the channel and the data symbols. The resulting \ac{BCD} algorithm, summarized in Algorithm~\ref{alg:gradient-descent}, iteratively mitigates grid mismatch effects by refining the estimated channel parameters and the associated data symbols, converging to a stationary point with improved estimation accuracy. Since the off-grid refinement stage can be carried out independently for each user, we describe it for the $k$-th user in the following.

\begin{algorithm}[t!]
\small
\SetAlgoLined
\setlength{\AlCapSkip}{0.75em}
\caption{Off-grid \ac{B-CE-DD} via \ac{BCD}}
\label{alg:gradient-descent}

\KwInput{Effective received signals $\{\acute{\Y}_k\}_{k=1}^{K}$; maximum number of iterations $I_{\textnormal{max}}$ and threshold $\epsilon_{2}$;}\nllabel{alg2:input}

\textbf{Initialization: } Set $\{\hat\thetab_k^{(0)}, \hat\r_k^{(0)}, \hat{\gammab}_k^{(0)}, \hat{\deltab}^{(0)}_k\}_{k=1}^{K}$ to the estimated channel parameters and data vectors obtained from Algorithm~\ref{algo1}; set $i=0$\;\nllabel{algo2:init}
\For{$k = 1, \ldots, K$}{%
    \nllabel{alg2:for-k-start}%
    \Repeat{$i = I_{\textnormal{max}}$  \textbf{\textup{or}} $F_k(\hat\thetab^{(i)}_k,\hat\r^{(i)}_k, \hat\gammab^{(i)}_k, \hat\deltab^{(i)}_k)\leq \epsilon_2$}{%
        \nllabel{alg2:for-n-start}%
        Set $i=i+1$\;
        Update the estimated angles $\thetab^{(i)}_k$ as in \eqref{update theta}\;\nllabel{alg2:update-theta}
        Update the estimated distances $\r^{(i)}_k$ as in \eqref{update r}\;\nllabel{alg2:update-r}
        Update the estimated small-scale fading coefficients $\hat{\gammab}_k^{(i)}$ as in
        \eqref{update G2}\;\nllabel{alg2:update-g}
        Update the estimated data symbols $\hat{\deltab}^{(i)}_k$ as in \eqref{update D2}\;\nllabel{alg2:update-gd}}\nllabel{alg2:end repeat}
    Set $\hat\h_k = \widetilde\W(\hat{\thetab}^{(i)}_k,\hat{\r}^{(i)}_k)\hat{\gammab}_k^{(i)}$
    and $\hat{\d}_k = \hat{\deltab}^{(i)}_k$\;\nllabel{alg2:set-h,d}
}

\KwOutput{$\hat\H = [\hat\h_1,\ldots,\hat\h_K]$ and $\hat{\D} = [\hat{\d}_1,\ldots,\hat{\d}_K]$.}\nllabel{alg2:output}
\end{algorithm}

At the $i$-th iteration, let $\hat{\thetab}_k^{(i)} = [\hat{\theta}_{k,1}^{(i)}, \ldots, \hat{\theta}_{k,\hat{L}}^{(i)}]^{\tran} \in \Real^{\hat{L}}$, $\hat{\r}_k^{(i)} = [\hat{r}_{k,1}^{(i)}, \ldots, \hat{r}_{k,\hat{L}}^{(i)}]^{\tran} \in \Real_{+}^{\hat{L}}$, and $\hat{\gammab}^{(i)}_k\in\Compl^{\hat{L}}$ represent the vectors containing the estimated angles, distances, and small-scale fading coefficients, respectively, associated with the $\hat{L}$ estimated propagation paths of the $k$-th user. Moreover, let $\hat{\deltab}^{(i)}\in\Compl^{S}$ denote the estimated data vector of the $k$-th user at the $i$-th iteration. Algorithm~\ref{alg:gradient-descent} is initialized with the output of Algorithm~\ref{algo1}. Specifically, we initialize $\hat{\thetab}_k^{(0)}$ and $\hat{\r}_k^{(0)}$ with the $\hat{L}$ angle and distance estimates obtained for the $k$-th user via \ac{B-OMP}, and set $\hat{\gammab}_k^{(0)} = \hat{\g}_k(\mathcal{S}_k^{(\hat{L})},:)$ and $\hat{\deltab}_k^{(0)} = \hat{\d}_k$, where $\mathcal{S}_k^{(\hat{L})}$ is the user-specific support set resulting from Algorithm~\ref{algo1} (see line~\ref{alg2:input}). These on-grid estimates serve as a suitable initialization for the off-grid refinement stage. Lastly, we define the effective polar-domain transformation matrix
\begin{align}\label{effective transformation matrix}
\widetilde\W(\hat{\thetab}_k^{(i)}, \hat{\r}_k^{(i)}) 
&= \big[ 
\b(\hat{\theta}_{k,1}^{(i)},\hat{r}_{k,1}^{(i)}), \ldots, 
\b(\hat{\theta}_{k,\hat{L}}^{(i)}, \hat{r}_{k,\hat{L}}^{(i)}) \big] \in \Compl^{N \times \hat{L}},
\end{align}
which comprises the steering vectors corresponding to the angle and distance estimates of the $k$-th user at the $i$-th iteration. This matrix is initialized as $\widetilde\W( \hat{\thetab}_k^{(0)}, \hat{\r}_k^{(0)}) = \W(:,\mathcal{S}_k^{(\hat{L})})$, consisting of the $\hat{L}$ columns of the polar-domain transformation matrix corresponding to the user-specific support set $\mathcal{S}_k^{(\hat{L})}$ obtained from Algorithm~\ref{algo1}.

Consider the effective received signal for the $k$-th user after removing the effect of the precoding matrix and the pilot symbol, defined as $\acute{\Y}_k = \Breve{\Y}_k(:,2{:}S{+}1) \in \Compl^{N \times S}$, with $\Breve{\Y}_k$ introduced in \eqref{Yk_block} and where we recall that the pilot symbol corresponds to the first column of $\Breve{\Y}_k$. For given channel and data estimates, the user-specific residual error between the effective received signal and its reconstruction defines the least-squares objective function
\begin{align}\label{cost function1}
    F_k(\hat\thetab_k,\hat\r_k, \hat\gammab_k, \hat\deltab_k)
    = \big\|\acute{\Y}_k
    - \widetilde\W(\hat\thetab_k, \hat\r_k)\hat\gammab_k\hat\deltab_k^\tran
    \big\|_\mathrm{F}^2.
\end{align}
Accordingly, the optimization problem for the $k$-th user is formulated as
\begin{align}\label{user_specific_obj1}
    \underset{\hat\thetab_k,
    \hat\r_k,  \hat\gammab_k, \hat\deltab_k}{\minimize}
    \; F_k(\hat\thetab_k,
    \hat\r_k,  \hat\gammab_k, \hat\deltab_k).
\end{align}
Since \eqref{cost function1} is not convex in $\hat\thetab_k$ and $\hat\r_k$ due to their dependence through the near-field steering vector in \eqref{near-field steering vector}, the problem in \eqref{user_specific_obj1} cannot be solved efficiently to global optimality. Hence, we adopt a \ac{BCD} approach that seeks a stationary point by sequentially updating each variable while keeping the others fixed.

At this stage, we note that the objective function in \eqref{cost function1} is quadratic in the estimated channel-data product $\hat{\gammab}_k \hat{\deltab}_k^\tran$, and minimizing it with respect to this product yields $\hat{\gammab}_k \hat{\deltab}_k^\tran = \widetilde\W(\hat{\thetab}_k, \hat{\r}_k)^\dag \acute{\Y}_k$. Substituting this expression back into \eqref{cost function1} removes the explicit dependence on $\hat{\gammab}_k$ and $\hat{\deltab}_k$, resulting in a reduced optimization problem that depends solely on the estimated angle and distance parameters. The reduced objective function is given by
\begin{align}\label{user_specific_obj_trace1}
\Phi_k(\hat\thetab_k, \hat\r_k) =  - \tr\big(\acute{\Y}_k^{\herm} \boldsymbol{\Psi}(\hat\thetab_k, \hat\r_k) \acute{\Y}_k\big),
\end{align}
where
\begin{align}\label{Psi_definition}
    \boldsymbol{\Psi}(\hat\thetab_k,\hat\r_k)
= \widetilde\W(\hat\thetab_k, \hat\r_k)
  \widetilde\W^{\dagger}(\hat\thetab_k, \hat\r_k)
\end{align}
denotes the orthogonal projection matrix onto the column space of \(\widetilde\W(\hat\thetab_k, \hat\r_k)\). The detailed derivations are provided in Appendix~\ref{Apx_1}. At the $i$-th iteration, the estimated angle and distance parameters are updated via a single \ac{GD} step applied to the reduced objective function in \eqref{user_specific_obj_trace1}, whereas the small-scale fading coefficients and the data vector are updated in closed form. The resulting \ac{BCD} updates are given as follows (see lines~\ref{alg2:update-theta}--\ref{alg2:update-gd}):
\begin{subequations}\label{eq:block-update}
\begin{align}
\hat{\thetab}_k^{(i)}
&\stackrel{\textrm{GD}}{\longleftarrow}
\Phi_k(\hat\thetab_k,\hat\r_k^{(i-1)}),
 \label{eq:block-update-a-nonconvex}\\
      \hat\r_k^{(i)}
&\stackrel{\textrm{GD}}{\longleftarrow}
\Phi_k(\hat\thetab_k^{(i)},\hat\r_k),
 \label{eq:block-update-b-nonconvex}\\  
   \hat\gammab^{(i)}_k 
   &= \argmin_{\hat\gammab_k} 
   F_k(\hat\thetab^{(i)}_k, \hat\r^{(i)}_k, \hat\gammab_k, \hat\deltab^{(i-1)}_k), \label{eq:block-update-c-convex}\\ 
   \hat\deltab^{(i)}_k 
   &= \argmin_{\hat\deltab_k} 
   F_k(\hat\thetab^{(i)}_k, \hat\r^{(i)}_k, \hat\gammab^{(i)}_k, \hat\deltab_k). \label{eq:block-update-d-convex}
\end{align}
\end{subequations}

The \ac{GD} step in \eqref{eq:block-update-a-nonconvex} is given by
\begin{align}\label{update theta}
    \hat{\thetab}_k^{(i)} = \hat{\thetab}^{(i-1)}_k - \eta_{\hat{\thetab}_k^{(i)}} \nabla_{\hat\thetab_k} \Phi_k(\hat{\thetab}_k, \hat{\r}^{(i-1)}_k)|_{\hat\thetab_k = \hat{\thetab}^{(i-1)}_k},
\end{align}
where $\eta_{\hat{\thetab}_k^{(i)}}$ denotes the step size and the gradient $\nabla_{\hat\thetab_k} \Phi_k(\hat{\thetab}_k, \hat{\r}_k^{(i-1)})$ is derived in Appendix~\ref{app:gradient_theta}. After refining the estimated angles, the effective polar-domain transformation matrix is updated as $\widetilde\W(\hat{\thetab}_k^{(i)}, \hat{\r}_k^{(i-1)})$, which is then used to refine the estimated distances. In line with the approach in~\cite{cui2022channel}, the \ac{GD} step for the estimated distances is performed with respect to the inverse vector $\frac{1}{\hat\r_k} = \big[\frac{1}{\hat r_{k,1}}, \ldots, \frac{1}{\hat r_{k,\hat{L}}}\big] \in \Real_{+}^{\hat{L}}$ rather than $\hat\r_k$ itself. This parametrization is consistent with the simulation setup, where the distance domain is uniformly sampled in the inverse of the distance, and has been observed to provide improved numerical stability. Accordingly, the \ac{GD} step in \eqref{eq:block-update-b-nonconvex} is carried out via
\begin{align}\label{update r}
    \frac{1}{\hat\r_k^{(i)}}= \frac{1}{\hat\r_k^{(i-1)} } - \eta_{\hat{\r}_k^{(i)}}\nabla_{ \frac{1}{\hat\r_k}}\Phi_k(\hat\thetab_k^{(i)}, \hat\r_k)|_{\hat\r_k = \hat\r^{(i-1)}_k},
\end{align}
where $\eta_{\hat{\r}_k^{(i)}}$ represents the step size and the gradient $\nabla_{\frac{1}{\hat\r_k}} \Phi_k(\hat\thetab_k^{(i)}, \hat\r_k)$ is derived in Appendix~\ref{app:gradient_r}. After refining the estimated distances, the effective polar-domain transformation matrix is again updated as $\widetilde\W(\hat{\thetab}_k^{(i)}, \hat{\r}_k^{(i)})$. The step sizes $\eta_{\hat{\thetab}_k^{(i)}}$ and $\eta_{\hat{\r}_k^{(i)}}$ are selected via backtracking line search to satisfy the descent condition and thus ensure smooth convergence to a stationary point of \eqref{cost function1}. After refining the estimated angle and distance parameters, the updates in \eqref{eq:block-update-c-convex} and \eqref{eq:block-update-d-convex} are carried out in closed form as
\begin{align}
\hat\gammab_k^{(i)} & =
\frac{1}{\|\hat{\deltab}_k^{(i-1)}\|_2^2}
\widetilde\W(\hat\thetab_k^{(i)},\hat\r_k^{(i)})^\dagger
\acute{\Y}_k(\hat{\deltab}_k^{(i-1)})^{\ast}, \label{update G2} \\
\hat{\deltab}_k^{(i)} & =
\frac{1}{\big\|\widetilde\W(\hat\thetab_k^{(i)},\hat\r_k^{(i)})
\hat\gammab_k^{(i)}\big\|_2^2}
\acute{\Y}_k^\tran
\big(\widetilde\W(\hat\thetab_k^{(i)},\hat\r_k^{(i)})
\hat\gammab_k^{(i)}\big)^{\ast},
\label{update D2}
\end{align}
respectively, which are precisely the least-squares solutions of \eqref{cost function1} at the $i$-th iteration with respect to $\hat\gammab_k$ and $\hat\deltab_k$ when the other variable is fixed.

This iterative procedure is repeated until a predefined termination criterion is met. Specifically, the algorithm terminates when one of the following conditions is satisfied (see line~\ref{alg2:end repeat}):
\begin{itemize}
\item The maximum number of \ac{BCD} iterations, denoted by $I_{\textnormal{max}}$, is reached.
\item The residual error becomes sufficiently small, i.e., for $F_k(\hat\thetab^{(i)}_k,\hat\r^{(i)}_k, \hat\gammab^{(i)}_k, \hat\deltab^{(i)}_k) \leq \epsilon_2$, where $\epsilon_2$ is a predefined threshold.  
\end{itemize}
Finally, for each user~$k$, the estimated channel vector is reconstructed as $\hat\h_k = \widetilde\W(\hat{\thetab}^{(i)}_k, \hat{\r}^{(i)}_k)\hat\gammab_k^{(i)}$, whereas the estimated data vector is given by $\hat{\d}_{k} = \hat\deltab_k^{(i)}$ (see line~\ref{alg2:set-h,d}). Stacking all the users yields $\hat\H = [\hat\h_1,\ldots,\hat\h_K]$ and $\hat{\mathbf{D}} = [\hat{\mathbf{d}}_1, \ldots, \hat{\mathbf{d}}_K]$.

The iterate sequence generated by the proposed \ac{BCD} algorithm converges to a stationary point of the objective function in \eqref{cost function1}. On the one hand, $\hat{\thetab}_k$ and $\hat\r_k$ are updated at each iteration via a single \ac{GD} step as in \eqref{update theta}--\eqref{update r}, where using an Armijo backtracking line search ensures that the objective is reduced. On the other hand, $\hat{\boldsymbol{\gamma}}_k$ and $\hat{\boldsymbol{\delta}}_k$ are updated in closed form according to \eqref{update G2}--\eqref{update D2}, which yield global minimizers of the objective function with respect to these variables. Hence, these steps generate an iterate sequence that is monotonically non-increasing. In addition, since the objective function in \eqref{cost function1} is a squared Frobenius norm, the generated iterate sequence is bounded below and converges to a non-negative limit. Under these conditions, existing convergence results for \ac{BCD} with inexact updates imply that every limit point of the generated iterate sequence satisfies the first-order stationarity conditions of the objective~\cite{Bon11,Cas13}. Consequently, the proposed \ac{BCD} algorithm converges to a stationary solution of the optimization problem in \eqref{user_specific_obj1}.

\subsection{Computational Complexity}\label{Complexity}

\begin{table}[t!]
\small
\centering
\begin{tabular}{|c|c|}
\hline
\textbf{Algorithm} & \textbf{Complexity order} \\
\hline
\hline
\ac{B-OMP}  & $\mathcal{O}(N T Q \hat{L})$ \\
\ac{BCD}    & $\mathcal{O}(N^{2} K S I_{\textnormal{max}})$ \\
\hline
\end{tabular}
\caption{Computational complexity of the proposed algorithms.}\label{tab:complexity}
\end{table}


In this section, we analyze the computational complexity of the proposed algorithms, with complexity orders summarized in Table~\ref{tab:complexity}. Throughout the analysis, we assume $Q > T \ge K(S+1)$ and $T > N > S > \hat{L}$: these assumptions reflect the considered system dimensions and are used to identify and retain only the dominant complexity terms.

For the \ac{B-OMP} algorithm, the computational complexity is mainly associated with lines~\ref{algo1:corr}, \ref{algo1:proj}, \ref{algo1:resid}, and~\ref{algo1:prephase} of Algorithm~\ref{algo1}.
\begin{itemize}
\item In line~\ref{algo1:corr}, the double matrix product in \eqref{correlation matrix} has per-iteration complexity that scales as $\mathcal{O} \big(\min \big( N T Q + K T (S+1) Q, N K T (S+1) + N K (S+1) Q\big)\big)$, which is upper bounded by $\mathcal{O} \big(\min (N T Q + T^{2} Q, N T^{2} + N T Q)\big) = \mathcal{O} (N T^{2} + N T Q)$. The dominant term is $\mathcal{O} (N T Q)$, which leads to a complexity scaling of $\mathcal{O}(N T Q \hat{L})$ over $\hat{L}$ iterations.
\item In line~\ref{algo1:proj}, the pseudoinverse in \eqref{least-square-algo1} scales as $\mathcal{O}(N i^{2})$ while the subsequent matrix multiplication scales as $\mathcal{O}\big(N (S+1) i\big)$ at iteration $i$. Summing over the users and iterations, the overall complexity is upper bounded by $\mathcal{O}(N K \hat{L}^{3} + N T \hat{L}^{2})$,\footnote{This is due to the relations $\sum_{i=1}^{\hat{L}} i = \frac{1}{2} \hat{L} (\hat{L} + 1)$ and $\sum_{i=1}^{\hat{L}} i^{2} = \frac{1}{6} \hat{L} (\hat{L} + 1) (2 \hat{L} + 1)$. \label{ftn:2}} which is dominated by $\mathcal{O}(N T \hat{L}^{2})$.
\item In line~\ref{algo1:resid}, the double matrix product in \eqref{residual_algo1} has per-iteration complexity that scales as $\mathcal{O} \big(\min \big(N K (S+1) i + N K T (S+1), K T (S+1) i + N K T i \big)\big)$ due to the block sparsity of $\hat{\boldsymbol{\Xi}}^{(i)}$, which is upper bounded by $\mathcal{O} \big(\min (N T i + N T^{2}, T^{2} i + N K T i)\big) = \mathcal{O} (T^{2} i + N K T i)$. The dominant term is $\mathcal{O} (T^{2} i)$, which leads to a complexity scaling of $\mathcal{O}(T^{2} \hat{L}^{2})$ over $\hat{L}$ iterations (see Footnote~\ref{ftn:2}).
\item In line~\ref{algo1:prephase}, the factorization scales as $\mathcal{O}\big((S+1)^{2} Q\big)$ when using \ac{SVD} or as $\mathcal{O}\big((S+1) Q I_{\textrm{PI}}\big)$ when using the power iteration method. Summing over the users and assuming the power iteration method, the complexity scaling is upper bounded by $\mathcal{O}(T Q I_{\textrm{PI}})$.
\end{itemize}
Finally, collecting the dominant terms, the total complexity of the \ac{B-OMP} algorithm scales as $\mathcal{O}(N T Q \hat{L} + N T \hat{L}^{2} + T^{2} \hat{L}^{2} + T Q I_{\textrm{PI}})$, which is dominated by $\mathcal{O}(N T Q \hat{L})$.

For the \ac{BCD} algorithm, the computational complexity is mainly associated with lines~\ref{alg2:update-theta}, \ref{alg2:update-r}, \ref{alg2:update-g}, and~\ref{alg2:update-gd} of Algorithm~\ref{alg:gradient-descent}.
\begin{itemize}
\item In lines~\ref{alg2:update-theta} and~\ref{alg2:update-r}, the \ac{GD} updates in \eqref{update theta} and \eqref{update r} require evaluating the gradients in \eqref{eq:grad_theta} and \eqref{eq:grad_r}, respectively (see Appendices~\ref{app:gradient_theta} and~\ref{app:gradient_r}). Each update involves a double matrix product scaling as $\mathcal{O}(N^{2} S + N S^{2})$, which is dominated by $\mathcal{O}(N^{2} S)$, and the derivatives in \eqref{eq:dPhi_theta} and \eqref{eq:dPhi_r}. The latter step involves three main operations: i) a double matrix product scaling as $\mathcal{O}\big(\min\big(N^2\hat{L}, N^2(N+\hat{L})\big)\big) = \mathcal{O}(N^2\hat{L} )$; ii) the computation of the projection matrix in \eqref{Psi_definition} scaling as $\mathcal{O}(N\hat{L}^{2} + N^{2}\hat{L})$, which is dominated by $\mathcal{O}(N^2\hat{L})$; and iii) the evaluation of the derivatives in \eqref{eq:dW_theta} and \eqref{eq:dW_r}, which involves only a single steering vector and scales as $\mathcal{O}(N)$. Summing over the users and iterations, the overall complexity of the \ac{GD} updates in lines~\ref{alg2:update-theta}--\ref{alg2:update-r} is $\mathcal{O}(N^{2}KSI_{\textnormal{max}} + N^{2}K\hat{L}I_{\textnormal{max}})$, which is dominated by $\mathcal{O}(N^{2} K S I_{\textnormal{max}})$.

\item In lines~\ref{alg2:update-g} and \ref{alg2:update-gd}, the double matrix products in \eqref{update G2} and \eqref{update D2} incur per-user, per-iteration complexity that scales as $\mathcal{O}\big(\min\big((N+1)S\hat{L},N(S+\hat{L})\big)\big) = \mathcal{O}\big(N(S+\hat{L})\big)$, which is dominated by $\mathcal{O}(N S)$. Summing over the users and iterations, the overall complexity scales as $\mathcal{O}(N K S I_{\textnormal{max}})$.
\end{itemize}
Finally, collecting the dominant terms, the total complexity of the \ac{BCD} algorithm scales as $\mathcal{O}(N^{2} K S I_{\textnormal{max}} + NKS I_{\textnormal{max}})$, which is dominated by $\mathcal{O}(N^{2} K S I_{\textnormal{max}})$.

\section{Numerical Results and Discussion} \label{simulation}

In this section, we evaluate the performance of the proposed \ac{B-CE-DD} framework against a baseline that employs \ac{ZF} beamforming at the \ac{BS} using channel estimates obtained with the \ac{OMP}-based method from~\cite{cui2022channel}, a widely adopted pilot-based approach for near-field \ac{XL-MIMO} channel estimation. To the best of our knowledge, no blind methods from the literature can be straightforwardly adapted to the considered scenario, and therefore suitable blind baselines for comparison are not available.

\subsection{Simulation Setup}

We consider a coherence interval of $T$ symbols, of which $S$ symbols are allocated to data transmission. The data symbols are drawn from an $M$-\ac{QAM} constellation. In the proposed \ac{B-CE-DD} framework, the transmitted signal of all the users are normalized as $\mathbb{E}\big[\|\x_k\|_2^2\big]=T$, $\forall k \in \{1, \ldots K\}$, leading to unit average transmit power per symbol. In the pilot-based baseline, the remaining $\tau = T-S$ symbols are reserved for pilot transmission. The pilot sequences are chosen as the first $K$ columns of a $\tau$-dimensional discrete Fourier transform matrix. Based on these orthogonal pilots, channel estimation is performed using the \ac{OMP}-based method from~\cite{cui2022channel}, and the resulting estimates are used to implement \ac{ZF} beamforming at the \ac{BS}. To ensure a fair comparison, the per-symbol average transmit power is kept constant across both schemes. Specifically, the pilot-based baseline assumes unit-power pilot symbols (i.e., total pilot energy of $T - S$) and normalized data symbols satisfying $\mathbb{E}\big[\|\d_k\|_2^2\big]=S$ (i.e., total data energy of $S$), yielding unit average transmit power per symbol.

We consider a \ac{ULA} with inter-antenna spacing $d = \frac{\lambda_\textrm{c}}{2}$, resulting in an array aperture of approximately $N \frac{\lambda_\textrm{c}}{2}$. The corresponding Fraunhofer distance, used to separate the near-field and far-field regions, is given by $R_{\textrm{Fraun}} \approx \frac{2}{\lambda_\textrm{c}} \big( N \frac{\lambda_\textrm{c}}{2} \big)^{2} = N^2 \frac{\lambda_\textrm{c}}{2}$. We consider a carrier frequency of $100$~GHz, corresponding to a wavelength of $\lambda_\textrm{c} = 3$~mm, $N \in \{128, 256\}$ antennas, and $K \in \{4, 6, 8\}$ users.

Unless otherwise stated, $L=6$ propagation paths are assumed for each user. The angle and distance corresponding to the $l$-th path of the $k$-th user are generated as $\theta_{k,l} \sim \mathcal{U} \big[-\frac{\pi}{4}, \frac{\pi}{4}\big]$ and $r_{k,l} \sim \mathcal{U} \big[ \frac{R_{\textrm{Fraun}}}{20}, \frac{2 R_{\textrm{Fraun}}}{3} \big]$, respectively. Moreover, the small-scale fading coefficient associated with the $l$-th path of the $k$-th user is generated as $g_{k,\ell} \sim \mathcal{CN}(0,1)$: with this choice, the expected channel power satisfies $\mathbb{E}\big[\|\h_k\|_2^2\big]=N$ according to \eqref{near-field channel}--\eqref{near-field steering vector}. In addition, the \ac{AWGN} variance is fixed to $\sigma^{2} = 1$ such that the per-antenna, per-symbol \ac{SNR} corresponds to $\rho$. We consider the \ac{SNR} range $\rho \in [-10, 10]$~dB. Such relatively low \ac{SNR} values are relevant for \ac{sub-THz} systems characterized by severe path loss and large bandwidths \cite{Atz25}.\footnote{As a representative example, an \ac{SNR} of $- 10$~dB at $100$~GHz can be achieved with bandwidth $B = 1$~GHz ($1\%$ of the carrier frequency), transmit power of $23$~dBm, noise figure $\nu = 8$~dB, and link distance of $67.3$~m, where the noise power in dB is given by $\sigma^{2} = -174 + 10 \log_{10} (B) + \nu$; in this setting, the Fraunhofer distance for $N = 256$ is $R_{\textrm{Fraun}} \approx 98.3$~m.}

The polar-domain transformation matrix $\mathbf{W}$ is constructed following~\cite{cui2022channel}. Specifically, the angular domain is uniformly sampled as $\bar{\theta}_{n} = \arcsin \big(\frac{2n-N+1}{N}\big)$, $\forall n \in \{0, \ldots, N -1 \}$. For each angular sample $\bar{\theta}_{n}$, a set of distance samples $\{\bar{r}_{n,v}\}_{v=1}^{V_{n}}$ is generated by uniformly sampling the distance domain in the inverse of the distance. The steering vectors corresponding to the resulting angle-distance pairs are then collected to form $\mathbf{W}$, where the total number of dictionary atoms is given by $Q=\sum_{n=0}^{N-1}V_{n}$.

\subsection{Results}

In the following, the curves marked with \ac{B-OMP} and \ac{B-OMP}+\ac{BCD} correspond to the proposed on-grid stage alone and the full \ac{B-CE-DD} framework (combining on-grid stage and off-grid refinement stage), respectively, whereas the considered baseline is referred to as \ac{OMP}+\ac{ZF}.


\begin{figure*}[t!]
\centering
\input{Results/figures/SER_SNR_group}
\caption{\ac{SER} versus $\rho$, with $K\in\{4,8\}$, $T=200$, $S=16$, and $M\in\{16,32,64\}$. Solid and dashed curves correspond to $N=128$ and $N=256$, respectively.}
\label{fig:SER_vs_SNR}
\end{figure*}
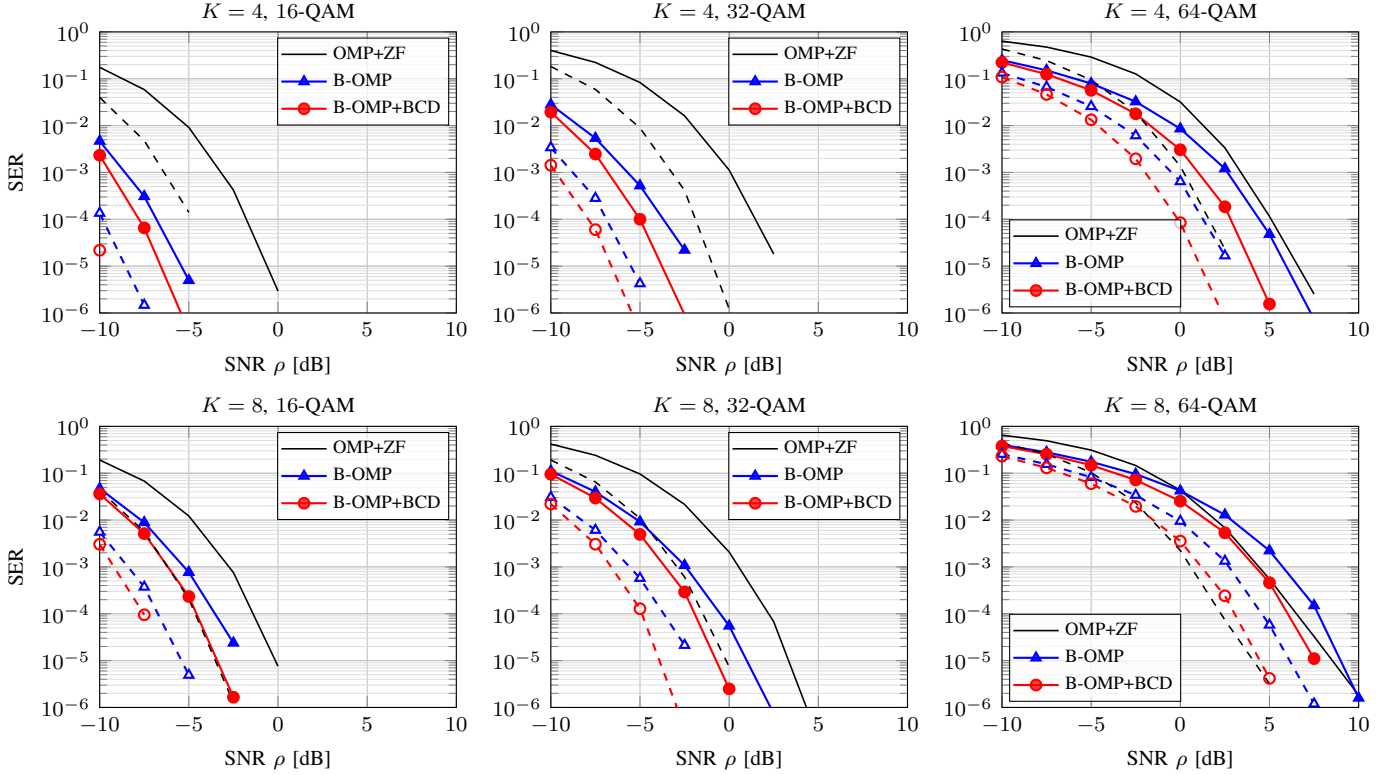

Fig.~\ref{fig:SER_vs_SNR} shows the \ac{SER} versus the \ac{SNR} $\rho$, with $K\in\{4,8\}$ users, coherence interval $T=200$ symbols, $S=16$ data symbols, and modulation order $M\in\{16,32,64\}$, where solid and dashed curves correspond to $N=128$ and $N=256$ antennas, respectively. As expected, the \ac{SER} for all the methods decreases with the \ac{SNR}, and the use of a larger array improves the performance due to the higher spatial resolution. The proposed \ac{B-CE-DD} framework provides significant gains over \ac{OMP}+\ac{ZF} in most considered settings, with the only exception occurring when both the number of users and modulation order are high. For $16$-\ac{QAM} and $32$-\ac{QAM}, \ac{B-OMP} alone achieves up to three orders of magnitude improvement in the observable \ac{SER} over \ac{OMP}+\ac{ZF}, while the addition of \ac{BCD} provides up to one further order of magnitude improvement, with less pronounced gains for $K=8$ due to the stronger multi-user interference. For $64$-\ac{QAM}, the proposed \ac{B-CE-DD} framework still performs well for $K=4$, showing that it can support high-order modulations under a moderate user load. However, for $K=8$, the denser constellation combined with increased multi-user interference make reliable data recovery more challenging, and \ac{OMP}+\ac{ZF} is superior over most of the \ac{SNR} range. Overall, these results confirm the benefit of the proposed \ac{B-CE-DD} framework and highlight the role of the off-grid refinement stage in mitigating the grid mismatch effects of the on-grid estimation.

\begin{figure}[t!]
\centering
\input{Results/figures/NMSE_vs_SER}
\caption{NMSE versus $\rho$, with $N=128$, $K = 4$, $T=200$, $S=16$, and $16$-\ac{QAM}.}
\label{fig:NMSE_vs_SNR}
\end{figure}
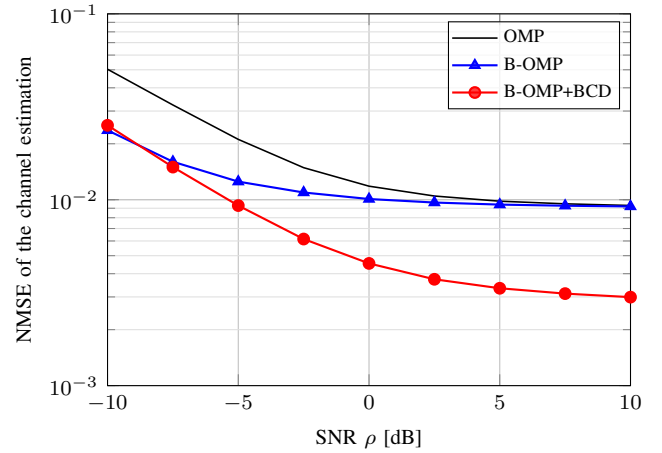

Although the primary objective of the proposed \ac{B-CE-DD} framework is data detection rather than channel estimation, it is still useful to assess the quality of the recovered channels. Fig.~\ref{fig:NMSE_vs_SNR} plots the \ac{NMSE} of the channel estimation versus the \ac{SNR} $\rho$, with $N =128$ antennas, $K = 4$ users, coherence interval $T=200$ symbols, $S=16$ data symbols, and $16$-\ac{QAM}. This corresponds to the setting in the top-left sub-figure in Fig.~\ref{fig:SER_vs_SNR}, whereas nearly indistinguishable curves are obtained for the other choices of $K$ and $N$. \ac{B-OMP} reduces the \ac{NMSE} by about $50$\% at low \ac{SNR} compared with the \ac{OMP}-based method from~\cite{cui2022channel}, while the benefit of applying \ac{BCD} on top of \ac{B-OMP} becomes more evident as the \ac{SNR} increases. These gains should not be viewed as the sole source of the improvements in data detection. Rather, the main advantage comes from jointly recovering the channel and transmitted data, which exploits the coupled structure of the received signal more efficiently than treating channel estimation and data detection as separate tasks.

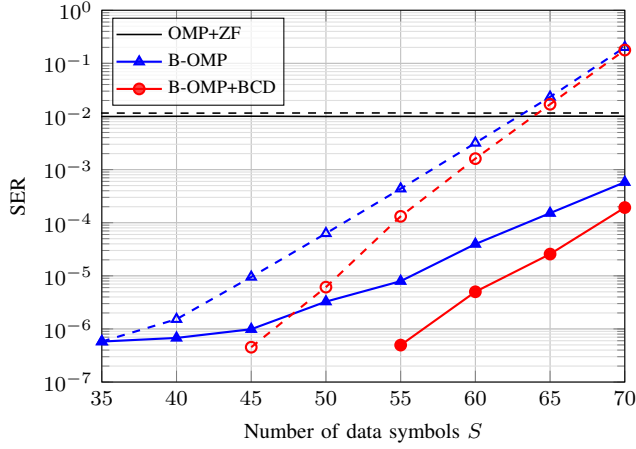
\begin{figure}[t!]
\centering
\input{Results/figures/SER_S_Diff_K}
\caption{\ac{SER} versus $S$, with $N=128$, $T=600$, $16$-\ac{QAM}, and $\rho=-5$~dB. Solid and dashed curves correspond to $K=6$ and $K=8$, respectively.}
\label{fig:SER_S_Diff_K}
\end{figure}

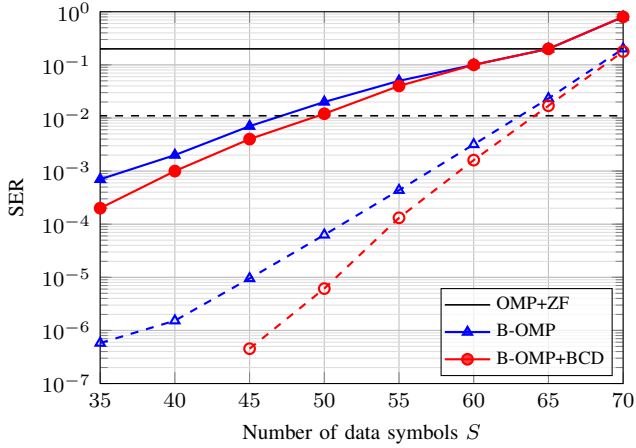
\begin{figure}[t!]
\centering
\input{Results/figures/SER_S_Diff_SNR}
\caption{\ac{SER} versus $S$, with $N=128$, $K=8$, $T=600$, and $16$-\ac{QAM}. Solid and dashed curves correspond to $\rho=-10$~dB, and $\rho=-5$~dB, respectively.}
\label{fig:SER_S_Diff_SNR}
\end{figure}

Fig.~\ref{fig:SER_S_Diff_K} and Fig.~\ref{fig:SER_S_Diff_SNR} illustrate the \ac{SER} versus the number of data symbols $S$, with $N=128$ antennas, coherence interval $T=600$ symbols, and $16$-\ac{QAM}. Fig.~\ref{fig:SER_S_Diff_K} considers a fixed \ac{SNR} $\rho=-5$~dB and shows the impact of the number of users, with solid and dashed curves corresponding to $K=6$ and $K=8$, respectively. Fig.~\ref{fig:SER_S_Diff_SNR} considers a fixed number of users $K=8$ and shows the impact of the \ac{SNR}, with solid and dashed curves corresponding to $\rho=-10$~dB, and $\rho=-5$~dB, respectively. Since the proposed \ac{B-CE-DD} framework requires the effective user-specific signals to be separable, only values of $S$ satisfying $T \geq K(S+1)$ are considered. The \ac{SER} resulting from \ac{B-OMP} and \ac{B-OMP}+\ac{BCD} increases as $S$ grows. On the other hand, the \ac{SER} achieved by \ac{OMP}+\ac{ZF} remains approximately constant over the considered range of $S$, since varying $S$ only slightly changes the pilot length $\tau$ under such a large $T$. Notably, the proposed \ac{B-CE-DD} framework provides significant gains over \ac{OMP}+\ac{ZF} when $S$ remains sufficiently below the limit imposed by $T \ge K(S+1)$. Moreover, increasing the number of users (and thus, the total number of data symbols to be recovered within the same coherence interval) and decreasing the \ac{SNR} not only degrades the \ac{SER} but also leads to a reduced gap between \ac{B-OMP} and \ac{B-OMP}+\ac{BCD}.

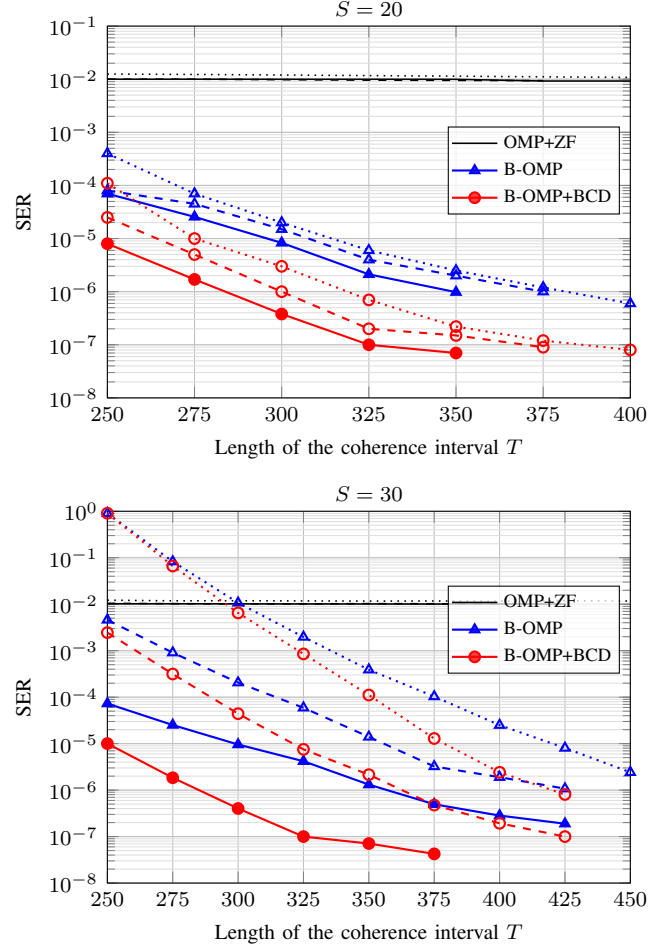
\begin{figure}[t!]
\centering
\input{Results/figures/SER_vs_T}
\caption{\ac{SER} versus $T$, with $N=128$, $S \in \{20,30\}$, $\rho=-5$~dB, and $16$-\ac{QAM}. Solid, dashed, and dotted curves correspond to $K = 4$, $K = 6$, and $K = 8$, respectively.}
\label{fig:SER_vs_T}
\end{figure}

Fig.~\ref{fig:SER_vs_T} shows the \ac{SER} versus the length of the coherence interval $T$, with $N=128$ antennas, $S \in \{20,30\}$ data symbols, \ac{SNR} $\rho=-5$~dB, and $16$-\ac{QAM}, where the solid, dashed, and dotted curves correspond to $K=4$, $K=6$, and $K=8$ users, respectively. For a fixed $S$, the \ac{SER} achieved by the proposed \ac{B-CE-DD} framework decreases as $T$ increases, since a larger coherence interval provides more observations for recovering the channel-data products, which improves both channel estimation and data detection. Conversely, for a fixed $T$, larger values of $K$ or $S$ lead to a higher total number of data symbols to be estimated from the same number of observations, thereby degrading the \ac{SER}.

\begin{figure}[t!]
\centering
\input{Results/figures/SER_vs_KS_over_T}
\caption{\ac{SER} versus $K\frac{S}{T}$, with $N=128$, $T=240$, $\rho=-5$ dB, and $16$-\ac{QAM}. Solid and dashed curves correspond to $S = 20$ (with $K = 12 x$) and $S = 30$ (with $K = 8 x$), respectively.}
\label{fig:SER_vs_K*S/T}
\end{figure}
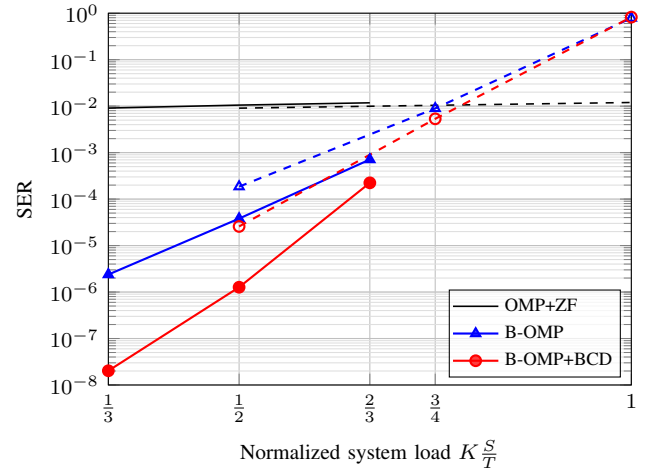

Fig.~\ref{fig:SER_vs_K*S/T} plots the \ac{SER} versus the normalized system load $K\frac{S}{T}$, with $N=128$ antennas, $K \in \{4, 6, 8\}$ users, coherence interval $T=240$ symbols, \ac{SNR} $\rho=-5$~dB, and $16$-\ac{QAM}, where the solid and dashed curves correspond to $S=20$ and $S=30$ data symbols, respectively. As expected, the \ac{SER} resulting from the proposed \ac{B-CE-DD} framework increases as $K\frac{S}{T}$ grows, since a higher load means that more data symbols and/or more users are accommodated within the same coherence interval. Specifically, \ac{B-OMP} performs approximately the same as \ac{OMP}+\ac{ZF} at $K\frac{S}{T} = \frac{3}{4}$, and exceeding this load consistently results in worse performance than the baseline. Interestingly, for the same load, serving more users with fewer data symbols per user leads to better performance than serving fewer users with more data symbols per user.

\begin{figure}[t!]
\centering
\input{Results/figures/SER_vs_L}
\caption{\ac{SER} versus $L$, with $N =128$, $K = 8$, $T=200$, $S=20$, $\rho=-10$ dB, and $16$-\ac{QAM}.}
\label{fig:SER_vs_L}
\end{figure}
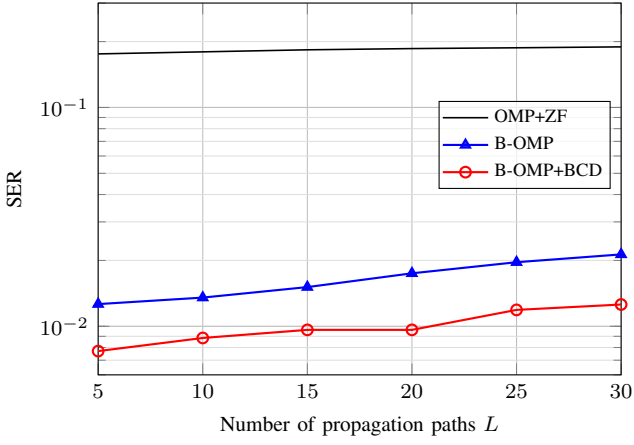

Lastly, Fig.~\ref{fig:SER_vs_L} illustrates the \ac{SER} versus the number of propagation paths $L$, with $N=128$ antennas, $K=8$ users, coherence interval $T=200$ symbols, $S=20$ data symbols, \ac{SNR} $\rho=-10$~dB, and $16$-\ac{QAM}. The \ac{SER} slightly increases as $L$ grows, since a larger number of paths corresponds to a less sparse channel in the polar domain, making the recovery of the channel-data products more challenging.

\section{Conclusions} \label{sec:concl}

In this paper, we investigated the problem of \acf{B-CE-DD} for uplink near-field \ac{XL-MIMO} systems. We formulated the considered problem as the recovery of user-specific rank-one channel-data products from a superimposed received signal by exploiting the polar-domain sparsity of near-field channels together with a low-dimensional data subspace model. Based on this formulation, we proposed a two-stage \ac{B-CE-DD} framework consisting of an on-grid algorithm, termed \acf{B-OMP}, and an off-grid refinement stage based on \acf{BCD}. The proposed \ac{B-OMP} algorithm iteratively identifies the dominant angle-distance components and estimates the corresponding channel-data products, while the proposed \ac{BCD} algorithm mitigates grid mismatch effects. Numerical results showed that the proposed \ac{B-CE-DD} framework achieves substantial performance gains relative to a pilot-based baseline employing \ac{ZF} beamforming, particularly at low \ac{SNR} and for low system load. Future work will focus on reducing the computational complexity of the proposed framework, accommodating a larger number of users through more scalable user separation and interference mitigation strategies, and extending the model to time-varying channels to avoid performing a full recovery procedure in every coherence interval.

\appendix

\subsection{Derivations of the Reduced Objective Function in \eqref{user_specific_obj_trace1}} \label{Apx_1}

Following similar steps as in~\cite{cui2022channel}, minimizing \eqref{cost function1} with respect to the channel-data product \(\hat\gammab_k \hat\deltab_k^\tran\) yields $\hat{\gammab}_k \hat{\deltab}_k^\tran = \widetilde\W(\hat{\thetab}_k, \hat{\r}_k)^\dag \acute{\Y}_k$. Substituting this expression back into \eqref{cost function1} and using the definition of the orthogonal projection matrix in \eqref{Psi_definition} results in
\begin{align}
\begin{split}
& F_k (\hat\thetab_k, \hat\r_k, \hat\gammab_k, \hat\deltab_k) \\
& = \big\| \acute{\Y}_k - \boldsymbol{\Psi}(\hat\thetab_k, \hat\r_k) \acute{\Y}_k \big\|_\mathrm{F}^2 \\
& = \tr\big(\acute{\Y}_k ^\herm\acute{\Y}_k - \acute{\Y}_k ^\herm \boldsymbol{\Psi}(\hat\thetab_k, \hat\r_k)\acute{\Y}_k - \acute{\Y}_k^\herm\boldsymbol{\Psi}^\herm(\hat\thetab_k, \hat\r_k) \acute{\Y}_k \\
& \phantom{=} \ + \acute{\Y}_k ^\herm\boldsymbol{\Psi}^\herm(\hat\thetab_k, \hat\r_k)\boldsymbol{\Psi}(\hat\thetab_k, \hat\r_k) \acute{\Y}_k \big) \\
& = \tr(\acute{\Y}_k \acute{\Y}_k ^\herm) - \tr\big(\acute{\Y}_k^\herm\boldsymbol{\Psi}(\hat\thetab_k, \hat\r_k)\acute{\Y}_k\big),
\end{split}
\end{align}
where the last steps follows the properties \(\boldsymbol{\Psi}^\herm(\hat\thetab_k, \hat\r_k) = \boldsymbol{\Psi}(\hat\thetab_k, \hat\r_k)\) and \(\boldsymbol{\Psi}^\herm(\hat\thetab_k, \hat\r_k)\boldsymbol{\Psi}(\hat\thetab_k, \hat\r_k) = \boldsymbol{\Psi}(\hat\thetab_k, \hat\r_k)\) of the orthogonal projection matrix. The reduced objective function in \eqref{user_specific_obj_trace1} is then obtained by removing the constant term \(\tr(\acute{\Y}_k ^\herm \acute{\Y}_k )\).

\subsection{Gradient of \eqref{user_specific_obj_trace1} with Respect to $\hat{\thetab}_k$} \label{app:gradient_theta}

Let \(\hat\theta_{k,l}\) denote the \(l\)-th entry of $\hat{\thetab}_k$. The gradient of \eqref{user_specific_obj_trace1} with respect to $\hat{\thetab}_k$ is given by
\begin{align}
\nabla_{\hat\thetab_k} \Phi_k(\hat{\thetab}_k, \hat{\r}_k) = \bigg[ \frac{\partial \Phi_k(\hat\thetab_k,\hat\r_k)}
{\partial \hat\theta_{k,1}}, \ldots, \frac{\partial \Phi_k(\hat\thetab_k,\hat\r_k)}
{\partial \hat\theta_{k,\hat{L}}} \bigg]^{\tran},
\end{align}
with
\begin{align}\label{eq:grad_theta}
\frac{\partial \Phi_k(\hat\thetab_k,\hat\r_k)}
{\partial \hat\theta_{k,l}}
=
-\tr\bigg(
\acute{\Y}_k^{\herm}
\frac{\partial \boldsymbol{\Psi}(\hat\thetab_k,\hat\r_k)}
{\partial \hat\theta_{k,l}}
\acute{\Y}_k
\bigg).
\end{align}
The derivative of the orthogonal projection matrix in \eqref{Psi_definition} with respect to \(\hat\theta_{k,l}\) is obtained as
\begin{align}\label{eq:dPhi_theta}
\begin{split}
& \frac{\partial \boldsymbol{\Psi}(\hat\thetab_k,\hat\r_k)}{\partial \hat\theta_{k,l}} \\
& = \big(\mathbf{I} - \boldsymbol{\Psi}(\hat\thetab_k,\hat\r_k)\big)\frac{\partial \widetilde{\W}(\hat\thetab_k,\hat\r_k)}{\partial \hat\theta_{k,l}}\widetilde{\W}(\hat\thetab_k,\hat\r_k)^{\dag} \\
& \phantom{=} \ + \big(\widetilde{\W}(\hat\thetab_k,\hat\r_k)^\dag\big)^{\herm}\frac{\partial \widetilde{\W}(\hat\thetab_k,\hat\r_k)^{\herm}}{\partial \hat\theta_{k,l}}\big(\mathbf{I} - \boldsymbol{\Psi}(\hat\thetab_k,\hat\r_k)\big) \\
& = 2 \Re \bigg[\big(\mathbf{I} - \boldsymbol{\Psi}(\hat\thetab_k,\hat\r_k)\big)\frac{\partial \widetilde{\W}(\hat\thetab_k,\hat\r_k)}{\partial \hat\theta_{k,l}}\widetilde{\W}(\hat\thetab_k,\hat\r_k)^{\dag} \bigg].
\end{split}
\end{align}
Based on \eqref{effective transformation matrix}, the derivative of $\widetilde{\W}(\hat\thetab_{k},\hat \r_{k})$ with respect to $\hat\theta_{k,l}$ only affects its \(l\)-th column, yielding
\begin{align}\label{eq:dW_theta}
\frac{\partial \widetilde{\W}(\hat\thetab_{k},\hat \r_{k})}{\partial \hat\theta_{k,l}}
=
\bigg[
\mathbf 0,\ldots,
\frac{\partial \b(\hat\theta_{k,l},\hat r_{k,l})}
{\partial \hat\theta_{k,l}},
\ldots,\mathbf 0
\bigg],
\end{align}
where the $n$-th entry of $\frac{\partial \b(\hat\theta_{k,l},\hat r_{k,l})}{\partial \hat\theta_{k,l}}$, for $n \in \{0,\ldots,N-1\}$, is given by
\begin{align}\label{dervative of b wrt theta}
j \frac{2\pi}{\lambda_\textrm{c}} \frac{\hat r_{k,l} nd\cos\hat\theta_{k,l}}{\hat r_{k,l}^{(n)}}e^{-j\frac{2 \pi}{\lambda_\textrm{c} }(\hat r^{(n)}_{k,l}-\hat r_{k,l})}.
\end{align}

\subsection{Gradient of \eqref{user_specific_obj_trace1} with Respect to $\frac{1}{\hat{\r}_k}$} \label{app:gradient_r}

Let $\hat r_{k,l}$ denote the $l$-th entry of $\hat{\r}_k$. The gradient of \eqref{user_specific_obj_trace1} with respect to $\frac{1}{\hat{\r}_k}$ is given by
\begin{align}
\nabla_{\frac{1}{\hat{\r}_k}} \Phi_k(\hat{\thetab}_k, \hat{\r}_k) = \bigg[ \frac{\partial \Phi_k(\hat\thetab_k,\hat\r_k)}
{\partial \frac{1}{\hat{r}_{k,1}}}, \ldots, \frac{\partial \Phi_k(\hat\thetab_k,\hat\r_k)}
{\partial \frac{1}{\hat{r}_{k,\hat{L}}}} \bigg]^{\tran},
\end{align}
with
\begin{align}\label{eq:grad_r}
\frac{\partial \Phi_k(\hat\thetab_k,\hat\r_k)}
     {\partial \frac{1}{\hat r_{k,l}}}
=
-\tr\bigg(
\acute{\Y}_k^{\herm}
\frac{\partial \boldsymbol{\Psi}(\hat\thetab_k,\hat\r_k)}
     {\partial \frac{1}{\hat r_{k,l}}}
\acute{\Y}_k
\bigg).
\end{align}
Similar to \eqref{eq:dPhi_theta}, the derivative of the orthogonal projection matrix in \eqref{Psi_definition} with respect to $\frac{1}{\hat r_{k,l}}$ is obtained as
\begin{align}\label{eq:dPhi_r}
\begin{split}
& \frac{\partial \boldsymbol{\Psi}(\hat\thetab_k,\hat\r_k)}{\partial \frac{1}{\hat r_{k,l}}} \\
& = \big(\mathbf{I} - \boldsymbol{\Psi}(\hat\thetab_k,\hat\r_k)\big)\frac{\partial \widetilde{\W}(\hat\thetab_k,\hat\r_k)}{\partial \frac{1}{\hat r_{k,l}}}\widetilde{\W}(\hat\thetab_k,\hat\r_k)^{\dag} \\
& \phantom{=} \ + \big(\widetilde{\W}(\hat\thetab_k,\hat\r_k)^\dag\big)^{\herm}\frac{\partial \widetilde{\W}(\hat\thetab_k,\hat\r_k)^{\herm}}{\partial \frac{1}{\hat r_{k,l}}}\big(\mathbf{I} - \boldsymbol{\Psi}(\hat\thetab_k,\hat\r_k)\big) \\
& = 2 \Re \bigg[\big(\mathbf{I} - \boldsymbol{\Psi}(\hat\thetab_k,\hat\r_k)\big)\frac{\partial \widetilde{\W}(\hat\thetab_k,\hat\r_k)}{\partial \frac{1}{\hat r_{k,l}}}\widetilde{\W}(\hat\thetab_k,\hat\r_k)^{\dag} \bigg].
\end{split}
\end{align}
Similar to \eqref{eq:dW_theta}, the derivative of $\widetilde{\W}(\hat\thetab_{k},\hat \r_{k})$ with respect to $\frac{1}{\hat r_{k,l}}$ is given by
\begin{align}\label{eq:dW_r}
\frac{\partial \widetilde{\W}(\hat\thetab_{k},\hat \r_{k})}
     {\partial \frac{1}{\hat r_{k,l}}}
=
\bigg[
\mathbf 0,\ldots,
\frac{\partial \b(\hat\theta_{k,l},\hat r_{k,l})}
     {\partial \frac{1}{\hat r_{k,l}}},
\ldots,\mathbf 0
\bigg],
\end{align}
where the $n$-th entry of
$\frac{\partial \b(\hat\theta_{k,l},\hat r_{k,l})}
        {\partial \frac{1}{\hat r_{k,l}}}$,
for $n\in\{0,\ldots,N-1\}$, is given by
\begin{align}\label{eq:derivative-b-invr}
j \frac{2\pi}{\lambda_\textrm{c}}
\hat r_{k,l}^{2}
\bigg(
\frac{\hat r_{k,l}-n d\sin\hat\theta_{k,l}}{\hat r_{k,l}^{(n)}} - 1
\bigg)
e^{-j\frac{2\pi}{\lambda_\textrm{c}}(\hat r_{k,l}^{(n)}-\hat r_{k,l})} .
\end{align}

\addcontentsline{toc}{chapter}{References}
\bibliographystyle{IEEEtran}
\bibliography{refs_abbr,refs}

\end{document}

%% file: Definitions.tex
\renewcommand{\a}{\mathbf{a}}
\renewcommand{\b}{\mathbf{b}}

\renewcommand{\d}{\mathbf{d}}

\newcommand{\g}{\mathbf{g}}
\newcommand{\h}{\mathbf{h}}

\renewcommand{\r}{\mathbf{r}}

\renewcommand{\u}{\mathbf{u}}
\renewcommand{\v}{\mathbf{v}}

\newcommand{\x}{\mathbf{x}}

\newcommand{\1}{\mathbf{1}}

\newcommand{\A}{\mathbf{A}}

\newcommand{\C}{\mathbf{C}}
\newcommand{\D}{\mathbf{D}}

\newcommand{\G}{\mathbf{G}}
\renewcommand{\H}{\mathbf{H}}
\newcommand{\I}{\mathbf{I}}

\newcommand{\R}{\mathbf{R}}

\newcommand{\U}{\mathbf{U}}
\newcommand{\V}{\mathbf{V}}
\newcommand{\W}{\mathbf{W}}

\newcommand{\Y}{\mathbf{Y}}
\newcommand{\Z}{\mathbf{Z}}

\newcommand{\gammab}{\boldsymbol{\gamma}}
\newcommand{\deltab}{\boldsymbol{\delta}}

\newcommand{\thetab}{\boldsymbol{\theta}}

\newcommand{\setI}{\mathcal{I}}

\newcommand{\Compl}{\mbox{$\mathbb{C}$}}
\newcommand{\Real}{\mbox{$\mathbb{R}$}}

\newcommand{\argmax}{\operatornamewithlimits{argmax}}
\newcommand{\argmin}{\operatornamewithlimits{argmin}}
\newcommand{\blkdiag}{\mathrm{blkdiag}}

\newcommand{\herm}{\mathrm{H}}

\newcommand{\minimize}{\mathrm{minimize}}

\renewcommand{\Re}{\mathrm{Re}}

\newcommand{\tr}{\mathrm{tr}}
\newcommand{\tran}{\mathrm{T}}

\definecolor{oulu_blue}{HTML}{23408F}
\definecolor{oulu_green}{HTML}{39B54A}

\definecolor{red}{rgb}{1,0,0}
\definecolor{red_magenta}{rgb}{1,0,0.5}
\definecolor{magenta}{rgb}{1,0,1}
\definecolor{blue_magenta}{rgb}{0.5,0,1}
\definecolor{blue}{rgb}{0,0,1}
\definecolor{blue_cyan}{rgb}{0,0.5,1}
\definecolor{cyan}{rgb}{0,1,1}
\definecolor{green_cyan}{rgb}{0,1,0.5}
\definecolor{green}{rgb}{0,1,0}
\definecolor{green_yellow}{rgb}{0.5,1,0}
\definecolor{yellow}{rgb}{1,1,0}
\definecolor{red_yellow}{rgb}{1,0.5,0}

%% file: Acronyms.tex
\begin{acronym}
\acro{AWGN}{additive white Gaussian noise}
\acro{BCD}{block-coordinate descent}
\acro{BS}{base station}
\acro{B-CE-DD}[\text{B-CE\&DD}]{blind channel estimation and data detection}
\acro{B-OMP}{blind orthogonal matching pursuit}
\acro{CSI}{channel state information}
\acro{GD}{gradient descent}
\acro{ISAC}{integrated sensing and communication}
\acro{LoS}{line-of-sight}
\acro{MIMO}{multiple-input multiple-output}
\acro{NLoS}{non-line-of-sight}
\acro{OMP}{orthogonal matching pursuit}
\acro{QAM}{quadrature amplitude modulation}
\acro{RIS}{reconfigurable intelligent surfaces}
\acro{SER}{symbol error rate}
\acro{NMSE}{normalized mean-square error}
\acro{SNR}{signal-to-noise ratio}
\acro{SVD}{singular value decomposition}
\acro{sub-THz}{sub-terahertz}
\acro{ULA}{uniform linear array}
\acro{XL-MIMO}{extremely large-scale MIMO}
\acro{ZF}{zero-forcing}
\end{acronym}

%% file: Results/figures/SER_SNR_group.tex
\begin{tikzpicture}
\begin{groupplot}[
group style={group size=3 by 2, horizontal sep=1.25cm, vertical sep=1.5cm},
width=6.3cm,
height=5.3cm,
xmin=-10, xmax=10,
ymin=1e-6, ymax=1e0,
ytick={1e-6,1e-5,1e-4,1e-3,1e-2,1e-1,1e0},
ymode=log,
grid=both,
major grid style={gray!50},
minor grid style={gray!25},
label style={font=\footnotesize},
ticklabel style={font=\footnotesize},
title style={font=\footnotesize, yshift=-2mm},
unbounded coords=jump,
every axis plot/.append style={line width=1.1pt},
]

\nextgroupplot[
title={$K=4$, $16$-QAM},
xlabel={SNR $\rho$ [dB]},
ylabel={SER},
legend style={
    at={(0.98,0.98)},
    anchor=north east,
    font=\scriptsize,
    fill opacity=.75,
    draw opacity=1,
    text opacity=1,
    cells={anchor=west},
    inner sep=1pt,
},
]

\addplot[black, semithick]
table [x=SNR, y=OMP_ZF, col sep=space] {Results/New_file_txt/SER_SNR_N128_K4_M16.txt};
\addlegendentry{OMP+ZF}
\addplot[blue, thick, mark=triangle*, mark options={solid, fill=blue}]
table [x=SNR, y=BOMP, col sep=space] {Results/New_file_txt/SER_SNR_N128_K4_M16.txt};
\addlegendentry{B-OMP}
\addplot[red, thick, mark=*, mark options={solid, fill=red}]
table [x=SNR, y=BCD, col sep=space] {Results/New_file_txt/SER_SNR_N128_K4_M16.txt};
\addlegendentry{B-OMP+BCD}

\addplot[black, dashed, semithick]
table [x=SNR, y=OMP_ZF, col sep=space] {Results/New_file_txt/SER_SNR_N256_K4_M16.txt};
\addplot[blue, thick, dashed, mark=triangle*, mark options={solid, fill=white}]
table [x=SNR, y=BOMP, col sep=space] {Results/New_file_txt/SER_SNR_N256_K4_M16.txt};
\addplot[red, thick, dashed, mark=o, mark options={solid, fill=white}]
table [x=SNR, y=BCD, col sep=space] {Results/New_file_txt/SER_SNR_N256_K4_M16.txt};

\nextgroupplot[
title={$K=4$, $32$-QAM},
xlabel={SNR $\rho$ [dB]},
legend style={
    at={(0.98,0.98)},
    anchor=north east,
    font=\scriptsize,
    fill opacity=.75,
    draw opacity=1,
    text opacity=1,
    cells={anchor=west},
    inner sep=1pt,
},
]

\addplot[black, semithick]
table [x=SNR, y=OMP_ZF, col sep=space] {Results/New_file_txt/SER_SNR_N128_K4_M32.txt};
\addlegendentry{OMP+ZF}
\addplot[blue, thick, mark=triangle*, mark options={solid, fill=blue}]
table [x=SNR, y=BOMP, col sep=space] {Results/New_file_txt/SER_SNR_N128_K4_M32.txt};
\addlegendentry{B-OMP}
\addplot[red, thick, mark=*, mark options={solid, fill=red}]
table [x=SNR, y=BCD, col sep=space] {Results/New_file_txt/SER_SNR_N128_K4_M32.txt};
\addlegendentry{B-OMP+BCD}

\addplot[black, dashed, semithick]
table [x=SNR, y=OMP_ZF, col sep=space] {Results/New_file_txt/SER_SNR_N256_K4_M32.txt};
\addplot[blue, thick, dashed, mark=triangle*, mark options={solid, fill=white}]
table [x=SNR, y=BOMP, col sep=space] {Results/New_file_txt/SER_SNR_N256_K4_M32.txt};
\addplot[red, thick, dashed, mark=o, mark options={solid, fill=white}]
table [x=SNR, y=BCD, col sep=space] {Results/New_file_txt/SER_SNR_N256_K4_M32.txt};

\nextgroupplot[
title={$K=4$, $64$-QAM},
xlabel={SNR $\rho$ [dB]},
legend style={
    at={(0.02,0.02)},
    anchor=south west,
    font=\scriptsize,
    fill opacity=.75,
    draw opacity=1,
    text opacity=1,
    cells={anchor=west},
    inner sep=1pt,
},
]

\addplot[black, semithick]
table [x=SNR, y=OMP_ZF, col sep=space] {Results/New_file_txt/SER_SNR_N128_K4_M64.txt};
\addlegendentry{OMP+ZF}
\addplot[blue, thick, mark=triangle*, mark options={solid, fill=blue}]
table [x=SNR, y=BOMP, col sep=space] {Results/New_file_txt/SER_SNR_N128_K4_M64.txt};
\addlegendentry{B-OMP}
\addplot[red, thick, mark=*, mark options={solid, fill=red}]
table [x=SNR, y=BCD, col sep=space] {Results/New_file_txt/SER_SNR_N128_K4_M64.txt};
\addlegendentry{B-OMP+BCD}

\addplot[black, dashed, semithick]
table [x=SNR, y=OMP_ZF, col sep=space] {Results/New_file_txt/SER_SNR_N256_K4_M64.txt};
\addplot[blue, thick, dashed, mark=triangle*, mark options={solid, fill=white}]
table [x=SNR, y=BOMP, col sep=space] {Results/New_file_txt/SER_SNR_N256_K4_M64.txt};
\addplot[red, thick, dashed, mark=o, mark options={solid, fill=white}]
table [x=SNR, y=BCD, col sep=space] {Results/New_file_txt/SER_SNR_N256_K4_M64.txt};

\nextgroupplot[
title={$K=8$, $16$-QAM},
ylabel={SER},
xlabel={SNR $\rho$ [dB]},
legend style={
    at={(0.98,0.98)},
    anchor=north east,
    font=\scriptsize,
    fill opacity=.75,
    draw opacity=1,
    text opacity=1,
    cells={anchor=west},
    inner sep=1pt,
},
]

\addplot[black, semithick]
table [x=SNR, y=OMP_ZF, col sep=space] {Results/New_file_txt/SER_SNR_N128_K8_M16.txt};
\addlegendentry{OMP+ZF}
\addplot[blue, thick, mark=triangle*, mark options={solid, fill=blue}]
table [x=SNR, y=BOMP, col sep=space] {Results/New_file_txt/SER_SNR_N128_K8_M16.txt};
\addlegendentry{B-OMP}
\addplot[red, thick, mark=*, mark options={solid, fill=red}]
table [x=SNR, y=BCD, col sep=space] {Results/New_file_txt/SER_SNR_N128_K8_M16.txt};
\addlegendentry{B-OMP+BCD}

\addplot[black, dashed, semithick]
table [x=SNR, y=OMP_ZF, col sep=space] {Results/New_file_txt/SER_SNR_N256_K8_M16.txt};
\addplot[blue, thick, dashed, mark=triangle*, mark options={solid, fill=white}]
table [x=SNR, y=BOMP, col sep=space] {Results/New_file_txt/SER_SNR_N256_K8_M16.txt};
\addplot[red, thick, dashed, mark=o, mark options={solid, fill=white}]
table [x=SNR, y=BCD, col sep=space] {Results/New_file_txt/SER_SNR_N256_K8_M16.txt};

\nextgroupplot[
title={$K=8$, $32$-QAM},
xlabel={SNR $\rho$ [dB]},
legend style={
    at={(0.98,0.98)},
    anchor=north east,
    font=\scriptsize,
    fill opacity=.75,
    draw opacity=1,
    text opacity=1,
    cells={anchor=west},
    inner sep=1pt,
},
]

\addplot[black, semithick]
table [x=SNR, y=OMP_ZF, col sep=space] {Results/New_file_txt/SER_SNR_N128_K8_M32.txt};
\addlegendentry{OMP+ZF}
\addplot[blue, thick, mark=triangle*, mark options={solid, fill=blue}]
table [x=SNR, y=BOMP, col sep=space] {Results/New_file_txt/SER_SNR_N128_K8_M32.txt};
\addlegendentry{B-OMP}
\addplot[red, thick, mark=*, mark options={solid, fill=red}]
table [x=SNR, y=BCD, col sep=space] {Results/New_file_txt/SER_SNR_N128_K8_M32.txt};
\addlegendentry{B-OMP+BCD}

\addplot[black, dashed, semithick]
table [x=SNR, y=OMP_ZF, col sep=space] {Results/New_file_txt/SER_SNR_N256_K8_M32.txt};
\addplot[blue, thick, dashed, mark=triangle*, mark options={solid, fill=white}]
table [x=SNR, y=BOMP, col sep=space] {Results/New_file_txt/SER_SNR_N256_K8_M32.txt};
\addplot[red, thick, dashed, mark=o, mark options={solid, fill=white}]
table [x=SNR, y=BCD, col sep=space] {Results/New_file_txt/SER_SNR_N256_K8_M32.txt};

\nextgroupplot[
title={$K=8$, $64$-QAM},
xlabel={SNR $\rho$ [dB]},
legend style={
    at={(0.02,0.02)},
    anchor=south west,
    font=\scriptsize,
    fill opacity=.75,
    draw opacity=1,
    text opacity=1,
    cells={anchor=west},
    inner sep=1pt,
},
]

\addplot[black, semithick]
table [x=SNR, y=OMP_ZF, col sep=space] {Results/New_file_txt/SER_SNR_N128_K8_M64.txt};
\addlegendentry{OMP+ZF}
\addplot[blue, thick, mark=triangle*, mark options={solid, fill=blue}]
table [x=SNR, y=BOMP, col sep=space] {Results/New_file_txt/SER_SNR_N128_K8_M64.txt};
\addlegendentry{B-OMP}
\addplot[red, thick, mark=*, mark options={solid, fill=red}]
table [x=SNR, y=BCD, col sep=space] {Results/New_file_txt/SER_SNR_N128_K8_M64.txt};
\addlegendentry{B-OMP+BCD}

\addplot[black, dashed, semithick]
table [x=SNR, y=OMP_ZF, col sep=space] {Results/New_file_txt/SER_SNR_N256_K8_M64.txt};
\addplot[blue, thick, dashed, mark=triangle*, mark options={solid, fill=white}]
table [x=SNR, y=BOMP, col sep=space] {Results/New_file_txt/SER_SNR_N256_K8_M64.txt};
\addplot[red, thick, dashed, mark=o, mark options={solid, fill=white}]
table [x=SNR, y=BCD, col sep=space] {Results/New_file_txt/SER_SNR_N256_K8_M64.txt};

\end{groupplot}
\end{tikzpicture}

%% file: Results/figures/NMSE_vs_SER.tex
\begin{tikzpicture}

\begin{axis}[
width=8.5cm,
height=6.5cm,
xmin=-10, xmax=10,
ymin=1e-3, ymax=1e-1,
ytick={1e-3,1e-2,1e-1},
ymode=log,
    xlabel={SNR $\rho$ [dB]},
    ylabel={NMSE of the channel estimation},
    xlabel near ticks,
    ylabel near ticks,
    label style={font=\footnotesize},
    ticklabel style={font=\footnotesize},
    title style={font=\footnotesize, yshift=-2mm},
    xtick={-10,-5,0,5,10},
    ytick={1e-3,1e-2,1e-1,1e0},
   grid=both,
    major grid style={gray!50},
    minor grid style={gray!25},
    legend style={
        at={(0.98,0.98)},
        anchor=north east,
        font=\scriptsize,
        fill opacity=.75,
        draw opacity=1,
        text opacity=1,
        cells={anchor=west},
        inner sep=1pt,
    },
    legend cell align=left,
    unbounded coords=discard
]

\addplot[black, semithick]
table [x=SNR, y=OMP_ZF, col sep=space] {Results/New_file_txt/NMSE_vs_SNR_K4_M16_N128.txt};
\addlegendentry{OMP}
\addplot[blue, thick, mark=triangle*, mark options={solid, fill=blue}]
table [x=SNR, y=BOMP, col sep=space] {Results/New_file_txt/NMSE_vs_SNR_K4_M16_N128.txt};
\addlegendentry{B-OMP}
\addplot[red, thick, mark=*, mark options={solid, fill=red}]
table [x=SNR, y=BCD, col sep=space] {Results/New_file_txt/NMSE_vs_SNR_K4_M16_N128.txt};
\addlegendentry{B-OMP+BCD}





\end{axis}
\end{tikzpicture}

%% file: Results/figures/SER_S_Diff_K.tex
\begin{tikzpicture}

\begin{axis}[
    width=8.5cm,
    height=6.5cm,
    xmin=35, xmax=70,
    ymin=1e-7, ymax=1e0,
    ymode=log,
    xlabel={Number of data symbols $S$},
    ylabel={SER},
    xlabel near ticks,
    ylabel near ticks,
    label style={font=\footnotesize},
    ticklabel style={font=\footnotesize},
    title style={font=\footnotesize, yshift=-2mm},
    xtick={20,25,30,35,40,45,50,55,60,65,70},
    ytick={1e-7,1e-6,1e-5,1e-4,1e-3,1e-2,1e-1,1e0},
    grid=both,
    major grid style={gray!50},
    minor grid style={gray!25},
    legend style={
        at={(0.02,0.98)},
        anchor=north west,
        font=\scriptsize,
        fill opacity=.75,
        draw opacity=1,
        text opacity=1,
        cells={anchor=west},
        inner sep=1pt,
    },
    legend cell align=left,
    unbounded coords=discard
]

\addplot[
    black,
    semithick
]
table [x=DL, y=PilotK6, col sep=space] {Results/New_file_txt/SER_S_Diff_K.txt};
\addlegendentry{OMP+ZF}
\addplot[
    blue,
    mark=triangle*,
    thick
]
table [x=DL, y=BOMPK6, col sep=space] {Results/New_file_txt/SER_S_Diff_K.txt};
\addlegendentry{B-OMP}

\addplot[
    red,
    mark=*,
    thick
]
table [x=DL, y=BCDK6, col sep=space] {Results/New_file_txt/SER_S_Diff_K.txt};
\addlegendentry{B-OMP+BCD}

\addplot[
    black,
    dashed,
    semithick
]
table [x=DL, y=PilotK8, col sep=space] {Results/New_file_txt/SER_S_Diff_K.txt};

\addplot[
    blue,
    dashed,
    mark=triangle,
    thick,
    mark options={solid}
]
table [x=DL, y=BOMPK8, col sep=space] {Results/New_file_txt/SER_S_Diff_K.txt};

\addplot[
    red,
    dashed,
    mark=o,
    thick,
    mark options={solid}
]
table [x=DL, y=BCDK8, col sep=space] {Results/New_file_txt/SER_S_Diff_K.txt};

\end{axis}
\end{tikzpicture}

%% file: Results/figures/SER_S_Diff_SNR.tex
\begin{tikzpicture}

\begin{axis}[
    width=8.5cm,
    height=6.5cm,
    xmin=35, xmax=70,
    ymin=1e-7, ymax=1e0,
    ymode=log,
    xlabel={Number of data symbols $S$},
    ylabel={SER},
    xlabel near ticks,
    ylabel near ticks,
    label style={font=\footnotesize},
    ticklabel style={font=\footnotesize},
    title style={font=\footnotesize, yshift=-2mm},
    xtick={20,25,30,35,40,45,50,55,60,65,70},
    ytick={1e-7,1e-6,1e-5,1e-4,1e-3,1e-2,1e-1,1e0},
    grid=both,
    major grid style={gray!50},
    minor grid style={gray!25},
    legend style={
        at={(0.98,0.02)},
        anchor=south east,
        font=\scriptsize,
        fill opacity=.75,
        draw opacity=1,
        text opacity=1,
        cells={anchor=west},
        inner sep=1pt,
    },
    legend cell align=left,
    unbounded coords=jump
]

\addplot[
    black,
    semithick,
]
table [x=DL, y=Pilotm10, col sep=space] {Results/New_file_txt/SER_S_Diff_SNR.txt};
\addlegendentry{OMP+ZF}

\addplot[
    blue,
    mark=triangle*,
    thick
]
table [x=DL, y=BOMPm10, col sep=space] {Results/New_file_txt/SER_S_Diff_SNR.txt};
\addlegendentry{B-OMP}

\addplot[
    red,
    mark=*,
    thick
]
table [x=DL, y=BCDm10, col sep=space] {Results/New_file_txt/SER_S_Diff_SNR.txt};
\addlegendentry{B-OMP+BCD}

\addplot[
    black,
    dashed,
    semithick
]
table [x=DL, y=Pilotm5, col sep=space] {Results/New_file_txt/SER_S_Diff_SNR.txt};

\addplot[
    blue,
    dashed,
    mark=triangle,
    thick,
    mark options={solid}
]
table [x=DL, y=BOMPm5, col sep=space] {Results/New_file_txt/SER_S_Diff_SNR.txt};

\addplot[
    red,
    dashed,
    mark=o,
    thick,
    mark options={solid}
]
table [x=DL, y=BCDm5, col sep=space] {Results/New_file_txt/SER_S_Diff_SNR.txt};

\end{axis}
\end{tikzpicture}

%% file: Results/figures/SER_vs_T.tex
\begin{tikzpicture}
\begin{groupplot}[
    group style={group size=1 by 2, vertical sep=1.5cm},
    width=8.5cm,
    height=6.5cm,
    ymode=log,
    grid=both,
    major grid style={gray!50},
    minor grid style={gray!25},
    label style={font=\footnotesize},
    ticklabel style={font=\footnotesize},
    title style={font=\footnotesize, yshift=-2mm},
    legend cell align=left,
    unbounded coords=jump,
    every axis plot/.append style={line width=1.1pt},
]

\nextgroupplot[
    title={$S=20$},
    xlabel={Length of the coherence interval $T$},
    ylabel={SER},
    xmin=250, xmax=400,
    xtick={250,275,...,400},
    ymin=1e-8, ymax=1e-1,
    ytick={1e-8,1e-7,1e-6,1e-5,1e-4,1e-3,1e-2,1e-1},
    legend style={
        at={(0.98,0.73)},
        anchor=north east,
        font=\scriptsize,
        fill opacity=.75,
        draw opacity=1,
        text opacity=1,
        cells={anchor=west},
        inner sep=1pt,
    },
]

\addplot[black, semithick]
table [x index=0, y index=1, col sep=tab] {Results/New_file_txt/SER_T_K4_S20.txt};
\addlegendentry{OMP+ZF}

\addplot[blue, thick, mark=triangle*, mark options={solid, fill=blue}]
table [x index=0, y index=2, col sep=tab] {Results/New_file_txt/SER_T_K4_S20.txt};
\addlegendentry{B-OMP}

\addplot[red, thick, mark=*, mark options={solid, fill=red}]
table [x index=0, y index=3, col sep=tab] {Results/New_file_txt/SER_T_K4_S20.txt};
\addlegendentry{B-OMP+BCD}

\addplot[black, semithick, dashed]
table [x index=0, y index=1, col sep=tab] {Results/New_file_txt/SER_T_K6_S20.txt};

\addplot[blue, thick, dashed, mark=triangle, mark options={solid}]
table [x index=0, y index=2, col sep=tab] {Results/New_file_txt/SER_T_K6_S20.txt};

\addplot[red, thick, dashed, mark=o, mark options={solid}]
table [x index=0, y index=3, col sep=tab] {Results/New_file_txt/SER_T_K6_S20.txt};

\addplot[black, semithick, dotted]
table [x index=0, y index=1, col sep=tab] {Results/New_file_txt/SER_T_K8_S20.txt};

\addplot[blue, thick, dotted, mark=triangle, mark options={solid}]
table [x index=0, y index=2, col sep=tab] {Results/New_file_txt/SER_T_K8_S20.txt};

\addplot[red, thick, dotted, mark=o, mark options={solid}]
table [x index=0, y index=3, col sep=tab] {Results/New_file_txt/SER_T_K8_S20.txt};

\nextgroupplot[
    title={$S=30$},
    xlabel={Length of the coherence interval $T$},
    ylabel={SER},
    xmin=250, xmax=450,
    xtick={250,275,...,450},
    ymin=1e-8, ymax=1e0,
    ytick={1e-8,1e-7,1e-6,1e-5,1e-4,1e-3,1e-2,1e-1,1e0},
    legend style={
        at={(0.98,0.80)},
        anchor=north east,
        font=\scriptsize,
        fill opacity=.75,
        draw opacity=1,
        text opacity=1,
        cells={anchor=west},
        inner sep=1pt,
    },  
]

\addplot[black, semithick]
table [x index=0, y index=1, col sep=tab] {Results/New_file_txt/SER_T_K4_S30.txt};
\addlegendentry{OMP+ZF}

\addplot[blue, thick, mark=triangle*, mark options={solid, fill=blue}]
table [x index=0, y index=2, col sep=tab] {Results/New_file_txt/SER_T_K4_S30.txt};
\addlegendentry{B-OMP}

\addplot[red, thick, mark=*, mark options={solid, fill=red}]
table [x index=0, y index=3, col sep=tab] {Results/New_file_txt/SER_T_K4_S30.txt};
\addlegendentry{B-OMP+BCD}

\addplot[black, semithick,dashed]
table [x index=0, y index=1, col sep=tab] {Results/New_file_txt/SER_T_K6_S30.txt};

\addplot[blue, thick,dashed, mark=triangle, mark options={solid}]
table [x index=0, y index=2, col sep=tab] {Results/New_file_txt/SER_T_K6_S30.txt};

\addplot[red, thick, dashed, mark=o, mark options={solid}]
table [x index=0, y index=3, col sep=tab] {Results/New_file_txt/SER_T_K6_S30.txt};

\addplot[black, semithick,dotted]
table [x index=0, y index=1, col sep=tab] {Results/New_file_txt/SER_T_K8_S30.txt};

\addplot[blue, thick,dotted, mark=triangle, mark options={solid}]
table [x index=0, y index=2, col sep=tab] {Results/New_file_txt/SER_T_K8_S30.txt};

\addplot[red, thick,dotted, mark=o, mark options={solid}]
table [x index=0, y index=3, col sep=tab] {Results/New_file_txt/SER_T_K8_S30.txt};

\end{groupplot}
\end{tikzpicture}

%% file: Results/figures/SER_vs_KS_over_T.tex
\begin{tikzpicture}

\begin{axis}[
    width=8.5cm,
    height=6.5cm,
    xmin=1/3, xmax=1,
    ymin=1e-8, ymax=1e0,
    ymode=log,
    xlabel={Normalized system load $K\frac{S}{T}$},
    ylabel={SER},
    xlabel near ticks,
    ylabel near ticks,
    label style={font=\footnotesize},
    ticklabel style={font=\footnotesize},
    title style={font=\footnotesize, yshift=-2mm},
    xtick={1/3,1/2,2/3,3/4,1},
    ytick={1e-8,1e-7,1e-6,1e-5,1e-4,1e-3,1e-2,1e-1,1e0},
    xticklabels={$\frac{1}{3}$,$\frac{1}{2}$,$\frac{2}{3}$,$\frac{3}{4}$,$1$},
    grid=both,
    major grid style={gray!50},
    minor grid style={gray!25},
    legend style={
        at={(0.98,0.02)},
        anchor= south east,
        font=\scriptsize,
        fill opacity=.75,
        draw opacity=1,
        text opacity=1,
        cells={anchor=west},
        inner sep=1pt,
    },
    legend cell align=left,
    unbounded coords=discard
]

\addplot[
    black,
    semithick
]
table [x=X, y=OMP, col sep=space] {Results/New_file_txt/SER_KS_over_T_S20.txt};
\addlegendentry{OMP+ZF}

\addplot[
    blue,
    mark=triangle*,
    thick
]
table [x=X, y=BOMP, col sep=space] {Results/New_file_txt/SER_KS_over_T_S20.txt};
\addlegendentry{B-OMP}

\addplot[
    red,
    mark=*,
    thick
]
table [x=X, y=BCD, col sep=space] {Results/New_file_txt/SER_KS_over_T_S20.txt};
\addlegendentry{B-OMP+BCD}

\addplot[
    black,
    dashed,
    semithick
]
table [x=X, y=OMP, col sep=space] {Results/New_file_txt/SER_KS_over_T_S30.txt};

\addplot[
    blue,
    dashed,
    mark=triangle,
    thick,
    mark options={solid, fill=white}
]
table [x=X, y=BOMP, col sep=space] {Results/New_file_txt/SER_KS_over_T_S30.txt};

\addplot[
    red,
    dashed,
    mark=o,
    thick,
    mark options={solid, fill=white}
]
table [x=X, y=BCD, col sep=space] {Results/New_file_txt/SER_KS_over_T_S30.txt};

\end{axis}

\end{tikzpicture}

%% file: Results/figures/SER_vs_L.tex
\begin{tikzpicture}

\begin{axis}[
    width=8.5cm,
    height=6.5cm,
    xmin=5, xmax=30,
    ymin=6e-3, ymax=3e-1,
    ymode=log,
    xlabel={Number of propagation paths $L$},
    ylabel={SER},
    xlabel near ticks,
    ylabel near ticks,
    label style={font=\footnotesize},
    ticklabel style={font=\footnotesize},
    title style={font=\footnotesize, yshift=-2mm},
    xtick={5,10,15,20,25,30},
    ytick={1e-3,1e-2,1e-1,1e0},
    grid=both,
    major grid style={gray!50},
    minor grid style={gray!25},
    legend style={
        at={(0.98,0.5)},
        anchor=south east,
        font=\scriptsize,
        fill opacity=.75,
        draw opacity=1,
        text opacity=1,
        cells={anchor=west},
        inner sep=1pt,
    },
    legend cell align=left,
    unbounded coords=jump
]

\addplot[black, semithick]
table[x index=0, y index=1, col sep=space]
{Results/New_file_txt/SER_vs_L.txt};
\addlegendentry{OMP+ZF}

\addplot[blue, thick, mark=triangle*]
table[x index=0, y index=2, col sep=space]
{Results/New_file_txt/SER_vs_L.txt};
\addlegendentry{B-OMP}

\addplot[red, thick, mark=o]
table[x index=0, y index=3, col sep=space]
{Results/New_file_txt/SER_vs_L.txt};
\addlegendentry{B-OMP+BCD}




\end{axis}
\end{tikzpicture}